\documentclass[lettersize,journal]{IEEEtran}
\IEEEoverridecommandlockouts

\usepackage{cite}
\usepackage{amsmath,amssymb,amsfonts}
\usepackage{algorithm}
\usepackage{color}

\usepackage{array}
\usepackage{stfloats}
\usepackage{url}
\usepackage{verbatim}
\usepackage{upgreek}
\usepackage{graphicx,color,overpic,psfrag}
\usepackage{textcomp}
\usepackage[T1]{fontenc}
\usepackage{mathrsfs}
\usepackage{float}
\graphicspath{{Figures/}}
\usepackage[table]{xcolor}
\usepackage[bookmarks=true, colorlinks, linkcolor=black, anchorcolor=black, citecolor=black]{hyperref}
\usepackage{latexsym}
\usepackage{bm}
\usepackage{cases}
\usepackage{fancyhdr}
\usepackage{setspace}
\usepackage{algpseudocode}
\usepackage{blkarray}
\usepackage{booktabs}
\usepackage{multirow}
\usepackage{dsfont}
\usepackage{tabularx}
\usepackage{amsfonts}
\usepackage{tikz}
\usepackage{longtable}

\makeatletter
\newenvironment{breakablealgorithm}
{
		\begin{center}
			\refstepcounter{algorithm}
			\hrule height.8pt depth0pt \kern2pt
			\renewcommand{\caption}[2][\relax]{
				{\raggedright\textbf{\ALG@name~\thealgorithm} ##2\par}%
				\ifx\relax##1\relax 
				\addcontentsline{loa}{algorithm}{\protect\numberline{\thealgorithm}##2}%
				\else 
				\addcontentsline{loa}{algorithm}{\protect\numberline{\thealgorithm}##1}%
				\fi
				\kern2pt\hrule\kern2pt
			}
		}{
		\kern2pt\hrule\relax
	\end{center}
}
\makeatother

\usepackage{lipsum}
\newtheorem{theorem}{Theorem}

	\DeclareMathAlphabet\mathbfcal{OMS}{cmsy}{b}{n}
	\ifCLASSOPTIONcompsoc
	\usepackage[caption=false,font=normalsize,labelfont=sf,textfont=sf]{subfig}
	\else
	\usepackage[caption=false,font=footnotesize]{subfig}
	\fi
	
	\allowdisplaybreaks[4]

	\def\BibTeX{{\rm B\kern-.05em{\sc i\kern-.025em b}\kern-.08em
			T\kern-.1667em\lower.7ex\hbox{E}\kern-.125emX}}
	
\begin{document}
		
		\title{Massive MIMO-OFDM Channel Acquisition with Time-Frequency Phase-Shifted Pilots}
		\author{Jinke Tang,~\IEEEmembership{Student~Member, IEEE,} Xiqi~Gao,~\IEEEmembership{Fellow, IEEE},  Li You,~\IEEEmembership{Senior~Member, IEEE,}\\ Ding Shi,~\IEEEmembership{Member, IEEE,} Jiyuan Yang,~\IEEEmembership{Student~Member, IEEE,} Xiang-Gen~Xia,~\IEEEmembership{Fellow, IEEE,}\\Xinwei Zhao, and Peigang Jiang
			

			\thanks{Part of this work was presented at the IEEE GLOBECOM 2024 \cite{Tang2412Time}.}
			\thanks{Jinke Tang, Xiqi Gao,  Li You, Ding Shi and Jiyuan Yang are with the National Mobile Communications Research Laboratory, Southeast University, Nanjing 210096, China, and also with Purple Mountain Laboratories, Nanjing 211111, China  (e-mail: jktang@seu.edu.cn,
				xqgao@seu.edu.cn, lyou@seu.edu.cn, shiding@seu.edu.cn, jyyang@seu.edu.cn).}
			\thanks{Xiang-Gen Xia is with the Department of Electrical and Computer Engineering, University of Delaware, Newark, DE 19716, USA (e-mail: xianggen@udel.edu).}
			\thanks{Xinwei Zhao and Peigang Jiang are with Huawei Technologies Co., Ltd., Shanghai  210206, China (e-mail:zhaoxinwei, jiangpeigang@huawei.com).}
		}

		\maketitle

		\begin{abstract}
			In this paper, we propose a channel acquisition approach with time-frequency phase-shifted pilots (TFPSPs) for massive multi-input multi-output orthogonal frequency division multiplexing (MIMO-OFDM) systems. We first present a triple-beam (TB) based channel tensor model, allowing for the representation of the space-frequency-time (SFT) domain channel as the product of beam matrices and the TB domain channel tensor. By leveraging the specific characteristics of TB domain channels, we develop TFPSPs, where distinct pilot signals are simultaneously transmitted in the frequency and time domains. Then, we present the optimal TFPSP design and provide the corresponding pilot scheduling algorithm. Further, we propose a tensor-based information geometry approach (IGA) to estimate the TB domain channel tensors. Leveraging the specific structure of beam matrices and the properties of TFPSPs, we propose a low-complexity implementation of the tensor-based IGA. We validate the efficiency of our proposed channel acquisition approach through extensive simulations. Simulation results demonstrate the superior performance of our approach. 
			The proposed approach can effectively suppress inter-UT interference with low complexity and limited pilot overhead, thereby enhancing channel estimation performance. Particularly in scenarios with a large number of UTs, the channel acquisition method outperforms existing approaches by reducing the normalized mean square error (NMSE) by more than 8 dB.
		
		\end{abstract}
		
		
		\section{Introduction}
		
		Massive multiple-input multiple-output (MIMO), combined with orthogonal frequency division multiplexing (OFDM), has become a core technique in wireless communications. This technique facilitates simultaneous service to a large number of user terminals (UTs) and thereby significantly enhances the system efficiency. It is pivotal in 5G communications, and will maintain its significance in future 6G communications with further increased numbers of antennas at the base stations (BSs) \cite{Tang2412Time,10054381,you2021towards,9237116,10158439}.
		The performance of the massive MIMO-OFDM system is closely related to the quality of channel state information (CSI) acquisition, commonly achieved through pilot-assisted techniques, where the transmitter sends pilots and the receiver estimates the CSI from the received signals. Therefore, pilot design and channel estimation have received significant attentions. A common technique is to transmit phase-shifted pilots in the frequency domain and then perform channel estimation using the specific channel characteristics, especially those in the delay domain \cite{training}. In recent years, this technique has undergone further refinement through its integration with the spatial-frequency (SF) beam-based channel model \cite{9439804,10026502,9910031}.
		
		Regarding pilot design, phase-shifted orthogonal pilots (PSOPs) have become widely adopted for channel acquisition due to their effectiveness in eliminating pilot interference within the same cell \cite{8654199}.  However, the specific format of PSOPs results in pilot overhead proportional to the number of UTs \cite{you2015pilot}. This overhead becomes excessively large as the number of UTs increases, creating a bottleneck for the system \cite{Marzetta2010non}. This issue is especially critical in high mobility scenarios, where pilots must be transmitted more frequently.
		
		To address the pilot overhead issue, adjustable phase-shifted pilots (APSPs) have been proposed, which enable the utilization of more available pilots without increasing pilot overhead \cite{7332961}. Although APSPs do not satisfy phase-shift orthogonality, they can achieve the same performance as PSOPs under certain conditions \cite{7332961, Chen2021apsp, 6951471}. However, it is important to note that, as the number of UTs in the system increases, these conditions for APSPs become increasingly difficult to satisfy  \cite{9042356}. For future 6G communications, where connection density and UT capacity may be 10 to 100 times higher than those in 5G [2], [5], the use of APSPs is likely to introduce significant pilot interference, which would degrade channel estimation performance. Consequently, there is a pressing need to develop new pilot design schemes that not only save pilot overhead but also effectively suppress inter-UT interference while adapting to the evolving requirements of next-generation systems.

		
		Pilot design schemes need to be combined with appropriate channel estimation methods, which are typically based on the  minimum mean square error (MMSE) criterion or the maximum a posteriori (MAP) criterion \cite{schreier2010statistical}.  However, in massive MIMO-OFDM systems, direct MMSE or MAP estimation is impractical due to the high computational complexity involved \cite{6415397}. Instead, researchers have extensively explored the sparsity characteristics of channels in the SF beam domain under the SF beam-based channel model, transforming the complex channel estimation problem into a sparse signal recovery issue \cite{9910031,9967937,10404161}. Consequently, statistical inference methods have been adopted in massive MIMO systems, significantly reducing the complexity. Various statistical inference methods, such as the belief propagation (BP) algorithm \cite{1459044}, the approximate message passing (AMP) algorithm \cite{donoho2009message}, and its generalized counterpart, the generalized AMP (GAMP) algorithm \cite{6033942}, have proven effective in approximating MMSE estimation. Additionally, statistical inference through Bethe free energy minimization has given rise to the expectation propagation (EP) algorithm and its variant, the expectation propagation variant (EPV) algorithm  \cite{9351786}. Recently, information geometry-based channel estimation has emerged as a valuable statistical inference approach by treating the region defined by the parameters of posterior probability distribution functions (PDFs) as a manifold \cite{9910031,RIGA,yang2024simplified2,amari2016information}. 
		
		Nonetheless, as massive MIMO-OFDM systems continue to evolve, the numbers of BS antennas and UTs in the system are expected to increase significantly, further increasing the difficulty of channel estimation. However, the inherent sparsity of the channel in the angular, delay, and Doppler domains  provides an opportunity to efficiently obtain the channel information for a large number of UTs within limited time-frequency resources \cite{Shen,9440710}. This underscores the necessity of developing new channel models that can exploit multi-domain sparsity. Moreover, channel estimation algorithms with lower computational complexity and faster convergence are needed in practice.


		In this paper, we present a novel  channel model and, building upon the model, we propose a comprehensive channel acquisition approach for massive MIMO-OFDM systems, where  pilot design, pilot scheduling and channel estimation are involved. The main contributions are summarized as follows:
		\begin{itemize}
			\item[$\bullet$]
			We present a triple-beam (TB) based channel tensor model for the massive MIMO-OFDM system, where the space-frequency-time (SFT) domain channel is represented by a product of beam matrices and the TB domain channel tensor. With the proposed model, channel sparsity in the angular, delay, and Doppler domains can be simultaneously reflected in the TB domain.
			\item[$\bullet$]
			With the properties of TB domain channels, we propose time-frequency phase-shifted pilots (TFPSPs). TFPSPs are transmitted in a quite different way from conventional pilots, which enable the system to accommodate a much larger number of UTs without increasing pilot overhead. We also derive the optimal condition of TFPSPs, under which the  mean squared error (MSE) of the channel estimation can be minimized. On this basis, we propose a TFPSP scheduling method, which simultaneously exploits the channel sparsity across triple dimensions, to achieve near-optimal performance.
			\item[$\bullet$]
			Further, we formulate the channel tensor estimation problem as a statistical inference problem and propose a tensor-based IGA to solve it.  With the specific structure of the beam matrices, we also simplify the tensor-based IGA to a more computationally efficient version,   which significantly reduces the overall computational complexity. Subsequently, we utilize the estimated results to perform channel prediction.
		\end{itemize}	
		
		The rest of the paper is organized as follows: In Section~\ref{section2}, we present the TB-based channel tensor model. In Section~\ref{section3},  we  design the TFPSPs and  propose the corresponding pilot scheduling algorithm. In Section~\ref{section4}, the tensor-based IGA for the estimation of the TB domain channel tensor is proposed, and an efficient implementation is provided. Simulation results are shown in Section~\ref{section5}, and the paper is concluded in Section~\ref{section6}.

		\emph{Notations:} Bold lowercase (uppercase, calligraphic) letters are used to denote column vectors (matrices, tensors). ${\bf{0}}$ denotes the all-zeros vector, matrix or tensor. ${{\bf{I}}_N}$ denotes the $N \times N$ identity matrix. Superscript ${( \cdot )^{\rm T}}$, ${( \cdot )^{\rm H}}$, ${( \cdot )^*}$ denote the transpose, conjugate-transpose, conjugate operations, respectively. We adopt ${\left[ {\bf{a}} \right]_i}$, ${\left[ {\bf{A}} \right]_{i,j}}$ and ${\left[ {\mathbfcal A} \right]_{i,j,k}}$ to denote the $i$-th element of the vector ${\bf{a}}$, the $(i,j)$-th element of ${\bf{A}}$ and the $(i,j,k)$-th element of the tensor ${\mathbfcal A}$, respectively. $\otimes$ and $\odot $ denote the Kronecker product and the Hadamard product, respectively. The operator ${\rm Diag}\{  \cdot \} $ denotes the diagonal matrix of the vector along its main diagonal, while ${\rm diag}\{  \cdot \} $ denotes extracting the diagonal elements as a vector. ${\rm tr}\{  \cdot \}$ denotes the matrix trace operation and ${\rm{vec}}\left\{  \cdot  \right\}$ denotes the vectorization operation. $ \underline{\succ} {\bf{0}}$ denotes semi-positive definite. 
		${\mathbb N}_{+}$ denotes the set of positive integers. ${\mathbb{E}}\{  \cdot \} $ denotes the expectation operation. We adopt $ {{{\cal A} _1}} \times  {{{\cal A} _2}}$ as the Cartesian product of the sets ${{\cal A} _1}$ and ${{\cal A} _2}$. The operator $\backslash $ represents the set subtraction operation. The expression $\left\lfloor x \right\rfloor $  denotes the largest integer smaller than $x$, while $\left\lceil x \right\rceil $ denotes the smallest integer larger than $x$. ${\left\langle x \right\rangle _N}$ denotes the modulo-$N$ operation. ${{\bf{F}}_N}$ denotes the $N\times N$ discrete Fourier transform (DFT) matrix and ${\bf{f}}_{N,a}^{}$ denotes its ${\left\langle a \right\rangle _N}$-th column. Besides, we define the following matrix 
		\begin{equation}\label{Gamma}
			{\bf{\Gamma }}_{M,m} \buildrel \Delta \over = \left[ {\begin{array}{*{20}{c}}
					{\bf{0}}&{{{\bf{I}}_{M - {{\left\langle m \right\rangle }_M}}}}\\
					{{{\bf{I}}_{{{\left\langle m \right\rangle }_M}}}}&{\bf{0}}
			\end{array}} \right].
		\end{equation}
		\section{System Model with Tensor Representation }\label{section2}
	
		In this section, we begin by introducing fundamental concepts of tensor algebra and outlining the configuration of a massive MIMO-OFDM system operating in time-division duplex (TDD) mode. Following this, we present the TB-based channel tensor model, which represents the channel in the SFT domain as a product of beam matrices and the TB domain channel tensor. Furthermore, we elaborate on the channel sparsity based on the proposed model, which will be leveraged for channel acquisition.
		\subsection{Preliminaries of Tensor Algebra} 
		
		To facilitate our analysis of multi-dimensional signals, we adopt tensor-based expressions, which directly preserve the original dimensionality of the model and make representations corresponding to different dimensions more intuitive. Accordingly, we provide some tensor definitions. Further details can be found in \cite{opentensor,brazell2013solving,liang2019tensor}.
		
		Considering ${\mathbfcal A} $ $\in$ $ {{\mathbb{C}}^{{I_1} \times {I_2} \times \cdots \times {I_M} \times {J_1} \times {J_2} \times \cdots \times {J_K}}}$, ${\mathbfcal A}$ is a \emph{square tensor} when $M=K$ and $I_m=J_m$ for $m=1,2, \cdots,M$. For a square tensor ${\mathbfcal A}$, the elements ${\left[ {\mathbfcal A} \right]}_{{{i_1},{i_2}, \cdots ,{i_M},{j_1},{j_2}, \cdots ,{j_k}}}$ satisfying $i_1=j_1$, $\cdots$, $i_M=j_M$ are called the \emph{pseudo-diagonal} elements of ${\mathbfcal A}$, and ${\mathbfcal A}$ can be regarded as a \emph{pseudo-diagonal tensor} if all elements  in ${\mathbfcal A}$ are $0$ except its pseudo-diagonal ones. A pseudo-diagonal tensor whose pseudo-diagonal elements are all equal to $1$ is referred to as an \emph{identity tensor}.  For instance, an ${I_1} \times  \cdots  \times {I_M} \times {I_1} \times  \cdots  \times {I_M}$ identity tensor is denoted by ${{\mathbfcal I}_{{I_1}, \cdots  , {I_M}}}$ where ${\left[ {{{\mathbfcal I}_{{I_1}, \cdots ,{I_M}}}} \right]_{{i_1}, \cdots ,{i_M},{i_1}, \cdots ,{i_M}}} = 1$ for all $i_1$, $\cdots$, $i_M$ and $0$ everywhere else.
	The \emph{conjugate} of ${\mathbfcal A}$ is denoted as ${{\mathbfcal A}^*}$, which is in the same size of ${\mathbfcal A}$ and $\left[ {{{\mathbfcal A}^*}} \right]_{{i_1}, \cdots ,{i_M},{j_1}, \cdots ,{j_K}}^{} = \left[ {\mathbfcal A} \right]_{{i_1}, \cdots ,{i_M},{j_1}, \cdots ,{j_K}}^*$. Meanwhile, we define the $M$-\emph{transpose} of ${\mathbfcal A}$ as ${\mathbfcal A}_{ M }^{\rm{T}} \in {{\mathbb{C}}^{{J_1} \times  \cdots  \times {J_K} \times {I_1} \times  \cdots  \times {I_M}}}$, where ${[{\mathbfcal A}_{M}^{\rm{T}}]_{{j_1},   \cdots, {j_K}, {i_1},  \cdots, {i_M}}}$ $ = \left[ {\mathbfcal A} \right]_{{i_1},  \cdots, {i_M}, {j_1}, \cdots, {j_K}}$, i.e., swapping its first $M$ modes with the remaining modes.  The $M$-\emph{Hermitian} can also be denoted as  ${\mathbfcal A}_{ M }^{\rm{H}}$, satisfying ${\mathbfcal A}_{ M }^{\rm{H}} = {\left( {{\mathbfcal A}_{ M }^{\rm{T}}} \right)^*}$.

	
	Given a tensor  ${\mathbfcal X} \in {\mathbb{C}}^{{I_1} \times {I_2} \times  \cdots  \times {I_m} \times  \cdots  \times {I_M}}$ and a matrix ${\bf{X}} \in {{\mathbb{C}}^{{J} \times {I_m}}}$ , the \emph{m-mode product} between ${\mathbfcal X}$ and ${\bf{X}}$ is denoted as ${\bf{X}}{ \times _m}{\mathbfcal X}$ $ \in$ $ {{\mathbb{C}}^{{I_1} \times  \cdots  \times {I_{m - 1}} \times {J} \times {I_{m + 1}} \times  \cdots  \times {I_M}}}$. Elementwise, we have    \vspace{-0.1cm}
	\begin{equation}\label{mmode}
		\!\!\!{\left[ {{\bf{X}}{ \times _m}{\mathbfcal X}} \right]_{{i_1}, \cdots ,{j}, \cdots ,{i_M}}} = \sum\limits_{{i_m} = 0}^{{I_m}-1} {{{\left[ {\bf{X}} \right]}_{{j},{i_m}}}{{\left[ {\mathbfcal X} \right]}_{{i_1}, \cdots ,{i_m}, \cdots ,{i_M}}}}.
	\end{equation}
	
	With $\!\!{\mathbfcal A}$$\quad\!\!\!\! \in$ $\quad\!\!\!\! {{\mathbb{C}}^{{I_1} \times {I_2} \times  \cdots  \times {I_M} \times {J_1} \times {J_2} \times  \cdots  \times {J_K}}}$ and ${\mathbfcal B}$ $\quad\!\!\!\! \in $ $\quad\!\!\!\!{{\mathbb{C}}^{{J_1} \times {J_2} \times  \cdots  \times {J_K} \times {P_1} \times {P_2} \times  \cdots  \times {P_N}}}$, the \emph{Einstein} \emph{product} between ${\mathbfcal A}$ and ${\mathbfcal B}$ is denoted as ${{\mathbfcal A}{*_{K}}{\mathbfcal B}}$$ \in {{\mathbb{C}}^{{I_1} \times  \cdots  \times {I_M} \times {P_1} \times  \cdots  \times {P_N}}}$. Elementwise, we have 
	\begin{equation}\label{einstein}\resizebox{0.85\hsize}{!}
		{$\begin{array}{l}
				\!\!\!\!\!\!	{\left[ {{\mathbfcal A}{*_{K}}{\mathbfcal B}} \right]_{{i_1}, \cdots ,{i_M},{p_1}, \cdots ,{p_N}}}=\\\!\!\!\!\displaystyle\sum\limits_{{j_1} = 0}^{{J_1} - 1} { \cdots \displaystyle\sum\limits_{{j_K} = 0}^{{J_K} - 1} {{{\left[ {\mathbfcal A} \right]}_{{i_1}, \cdots ,{i_M},{j_1}, \cdots ,{j_K}}}} } {{\left[ {\mathbfcal B} \right]}_{{j_1}, \cdots ,{j_K},{p_1}, \cdots ,{p_N}}} .
			\end{array}$}
	\end{equation}For two matrices ${\bf{A}} \in {{\mathbb{C}}^{M \times N}}$ and ${\bf{B}} \in {{\mathbb{C}}^{N \times J}}$, the Einstein product ${\bf{A}} { * _{1}}{\bf{B}} $ is the standard matrix multiplication ${{\bf{AB}}}$. For ${\mathbfcal X} \in {{\mathbb{C}}^{{I_1} \times {I_2} \times  \cdots  \times {I_M}}}$ and ${\mathbfcal Y} \in {{\mathbb{C}}^{{J_1} \times {J_2} \times  \cdots  \times {J_K}}}$, with Einstein product we can define the norm of ${\mathbfcal X}$ as ${\left\| {\mathbfcal X} \right\|_2} = \sqrt {{\mathbfcal X}{ * _{M}}{{\mathbfcal X}^*}} $, and ${\left\| {\mathbfcal X} \right\|_0}$ denotes the number of non-zero elements in ${\mathbfcal X}$. The \emph{outer product} of  ${\mathbfcal X}$ and ${\mathbfcal Y}$ is denoted as ${\mathbfcal X} \circ {\mathbfcal Y}$$\in$${\mathbb C}^{{I_1}\times \cdots \times{I_M}\times{J_1}\times \cdots \times{J_K}}$, where
	\begin{equation}\label{outer}
		{\left[ {{\mathbfcal X} \circ {\mathbfcal Y}} \right]_{{i_1}, \cdots ,{i_M},{j_1}, \cdots ,{j_M}}} = {\left[ {\mathbfcal X} \right]_{{i_1}, \cdots ,{i_M}}}{\left[ {\mathbfcal Y} \right]_{{j_1}, \cdots ,{j_K}}}.
	\end{equation}

	With Einstein product, the inverse of the square tensor ${\mathbfcal Z} \in {{\mathbb{C}}^{{I_1} \times {I_2} \times  \cdots  \times {I_M} \times {I_1} \times {I_2} \times  \cdots  \times {I_M}}}$ can be denoted as ${{\mathbfcal Z}^{ - 1}}$ satisfying ${\mathbfcal Z}{ * _M}{{\mathbfcal Z}^{ - 1}} = {{\mathbfcal Z}^{ - 1}}{ * _M}{\mathbfcal Z} = {{\mathbfcal I}_{{I_1},{I_2}, \cdots ,{I_M}}}$, which is the same as the matrix inverse. For any tensor ${\mathbfcal A} \in {{\mathbb{C}}^{{I_1} \times {I_2} \times  \cdots  \times {I_M} \times {J_1} \times {J_2} \times  \cdots  \times {J_K}}}$,  the pseudo-inverse of ${\mathbfcal A}$, i.e., ${{{\mathbfcal A}^\dag }}$, is in the same size as ${{\mathbfcal A}}$, satisfying ${\left[ {{{\mathbfcal A}^\dag }} \right]_{{i_1}, \cdots ,{i_M},{j_1}, \cdots ,{j_K}}} = 1/{\left[ {\mathbfcal A} \right]_{{i_1}, \cdots ,{i_M},{j_1}, \cdots ,{j_K}}}$ once ${\left[ {\mathbfcal A} \right]_{{i_1}, \cdots ,{i_M},{j_1}, \cdots ,{j_K}}} \ne 0$ and the other elements are $0$. Besides, the \emph{trace} of a square tensor refers to the sum of its pseudo-diagonal elements, and for tensors ${\mathbfcal A} \in {{\mathbb{C}}^{{I_1} \times {I_2} \times  \cdots  \times {I_M} \times {J_1} \times {J_2} \times  \cdots  \times {J_K}}}$ and ${\mathbfcal B} \in {{\mathbb{C}}^{{J_1} \times {J_2} \times  \cdots  \times {J_K} \times {I_1} \times {I_2} \times  \cdots  \times {I_M}}}$, we have  ${\rm{tr}}\left\{ {{\mathbfcal A}{ * _K}{\mathbfcal B}} \right\}$ $ = $ ${\rm{tr}}\left\{ {{\mathbfcal B}{ * _M}{\mathbfcal A}} \right\}$. Actually, these tensor-based definitions can be regarded as the generalization of matrix algebra, and the properties of some matrix operations, such as matrix inversion and matrix multiplication, can be applied to the above tensor operations directly.
	\vspace{-0.25cm}		
	\subsection{System Configuration}

	We consider a single-cell TDD massive MIMO-OFDM system operating at the carrier frequency $f_{\rm c}$, in which the base station (BS) is equipped with an $M$ antenna uniform linear array (ULA) and serves $U$ single-antenna users terminals (UTs). With ${\lambda _{\rm{c}}}$ as the wavelength, the inter-antenna spacing $d$ of the ULA at the BS is set as $0.5{\lambda _{\rm{c}}}$. Let $ {\cal U} = \{ 0,1, \cdots ,U - 1\} $ denote the UT set. As for the OFDM modulation, ${N_{\rm c}}$ and ${N_{\rm g}}$ denote the number of subcarriers and the length of the cyclic prefix (CP), respectively. With the sampling interval length denoted by $T_{\rm s}$, the subcarrier spacing is $\Delta f = \frac{1}{{{N_{\rm c}}{T_{\rm s}}}}$, and the duration of each OFDM symbol is ${T_{{\rm{sym}}}} = ({N_{\rm c}} + {N_{\rm g}}){T_{\rm s}}$. Within each OFDM symbol, we use ${{{\cal K}}} = \{ {k_0},{k_0}+1, \cdots ,{k_0+{{K} - 1}}\} $ as the index set of the ${K}$ valid subcarriers used for training and data transmissions, and  ${{{\cal K}}}$ is a subset of $ \{ 0,1,\cdots,N_{\rm c}-1\} $.
	

	
	\begin{figure}[t!]
		\centering
		\includegraphics[width =210pt]{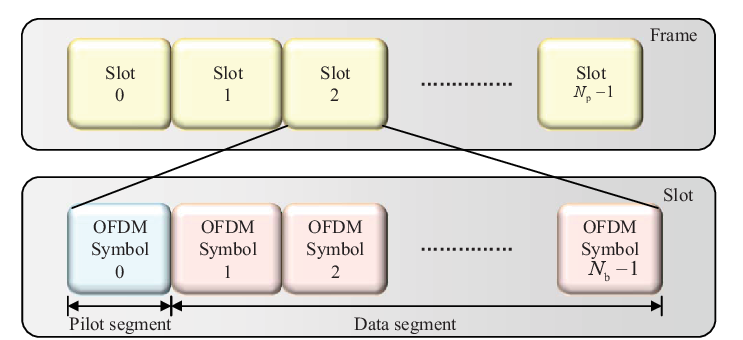}\label{frame0413}
		\caption{Frame structure for massive MIMO-OFDM transmissions} 
		\label{multimodel}
	\end{figure}
	
	
	
	The time resource is divided into frames, and each frame is divided into $N_{\rm{p}}$ slots, as shown in Fig. \ref{multimodel}. Each slot contains ${N_{\rm{b}}}$ OFDM symbols, where the first symbol in each slot is used for uplink (UL) pilot transmission and the other symbols are used for UL or downlink (DL) data transmissions. To acquire the CSI in each slot, we can combine the current slot with its previous ${\left( {{N_{\rm{p}}} - 1} \right)}$ slots to form a frame for channel estimation \cite{9967937}. 
	We assume that at the time slot $T$, the current frame for channel estimation is formed from the  $\left( {T - {N_{\rm{p}}} + 1} \right)$-th to the $T$-th slot. In this frame, with ${n_T} \buildrel \Delta \over =  ({T - {N_{\rm{p}}} + 1}) $, we can define ${{\cal P}_T} $ $\buildrel \Delta \over = $ $\left\{ {{n_T}{N_{\rm{b}}},} \right.$ ${{{n_T}}{N_{\rm{b}}} + 1,}$ ${ \cdots,}$ ${\left( {T + 1} \right){N_{\rm{b}}}}$ $\left. { - 1} \right\}$ as the OFDM symbol set, and a subset of ${{\cal P}_T} $ as the OFDM symbol set of pilot segments, i.e., ${{\cal P}_T^{\rm{p}}} \buildrel \Delta \over = \left\{ {{{n_T}}{N_{\rm{b}}},\left( {n_T}+1 \right){N_{\rm{b}}}, \cdots ,T{N_{\rm{b}}}} \right\}$. There are ${N_{\rm s}}=N_{\rm{b}} *{N_{\rm{p}}}$ OFDM symbols in this frame totally, and ${N_{\rm{p}}}$ pilot segments, each of which corresponds to one OFDM symbol. 

	\subsection{TB-based Channel Tensor Model}
	
	With the above system configuration, we propose the corresponding channel model. Considering a propagation environment where there are ${P_u}$ paths between the BS and the $u$-th UT.	The equivalent channel impulse response between the $m$-th antenna at the BS and the $u$-th UT can be expressed as \cite{868476,Shen,9440710}
	\begin{equation}\label{CIR}	\resizebox{0.888\hsize}{!}
		{${{\widetilde h}_{u,m}}\left( {t,\tau } \right)  = \sum\limits_{p = 0}^{{P_u} - 1} {{\alpha _{u,p}}{e^{\bar \jmath 2\pi {\nu _{u,p}}t}}{e^{ - \bar \jmath 2\pi m{\vartheta _{u,p}}}}\delta \left( {\tau  - {\tau _{u,p}}} \right)},$}
	\end{equation}
	where $u \in {\cal U} $, $m \in \{ 0,1, \cdots ,M - 1\} $. ${{\alpha _{u,p}}}$, ${{\nu _{u,p}}}$, ${{\vartheta _{u,p}}}$, ${{\tau _{u,p}}}$ represent the complex-valued channel gain, the Doppler frequency, the directional cosine and the delay of the $p$-th path, respectively. ${{\vartheta _{u,p}}}$ satisfies the relationship ${\vartheta _{u,p}} \buildrel \Delta \over = \frac{d}{{{\lambda _c}}}\cos {\theta _{u,p}} = 0.5\cos {\theta _{u,p}}$ with the incidence angle ${\theta _{u,p}}$. 
	
	Let ${{x_{u,k,n}}}$ denote the transmitted signal of the $u$-th UT at the $k$-th subcarrier of the $n$-th OFDM symbol, where $k \in {{\cal K}}$. We assume that the channel state remains constant within an OFDM symbol and varies symbol by symbol due to the Doppler effect. Based on (\ref{CIR}), the received signal on the $n$-th OFDM symbol at the $k$-th subcarrier can be written as
	\begin{align}\label{recy}
		\scalebox{0.95}{${y_{m,k,n}}$}& \scalebox{0.95}{$ = \sum\limits_{u = 0}^{U - 1} {h_{u,m,k,n}^{{\rm{SFT}}}{x_{u,k,n}}}  + \widetilde z$},
	\end{align}
	where ${\widetilde z}$ is the noise, and ${h_{u,m,k,n}^{\rm{SFT}}}$ is the channel response in the SFT domain between the $u$-th UT and the $m$-th antenna at the BS on the $k$-th subcarrier of the $n$-th OFDM symbol, which is given by \cite{9967937}
	\begin{equation}\label{gk}
		\resizebox{0.85\hsize}{!}
		{$\begin{array}{*{10}{l}}
				\!\!\!\!\!\!\!\!\!\!	h_{u,m,k,n}^{{\rm{SFT}}}
				\!\!\!\!\!\!\!\!&= \displaystyle\int {{\widetilde h}_{u,m}\left( {n{T_{\rm{sym}}},\tau } \right){e^{ - \bar \jmath 2\pi k\Delta f\tau }}d\tau } \\&=\!\!\displaystyle\sum\limits_{p = 0}^{{P_u} - 1} \!\!{{\alpha _{u,p}}{e^{\bar \jmath 2\pi {\nu _{u,p}}n{T_{\rm{sym}}}}}{{{e^{ - \bar \jmath 2\pi m{\vartheta _{u,p}}}}}} {e^{ - \bar \jmath 2\pi k\Delta f{\tau _{u,p}}}}}.
			\end{array}$}
	\end{equation}

	
	We use ${v_{{\rm{speed}}}}$ to denote the maximum UT speed in the system and ${\nu _{\max }} \buildrel \Delta \over = 2{v_{{\rm{speed}}}}/{\lambda _c}$. Assuming all the UTs are synchronized \cite{10066300}, we have ${\vartheta _{u,p}} \in \left[ { - 0.5,0.5} \right]$, ${\tau _{u,p}} \in [0,1/\Delta f]$ and ${\nu _{u,p}} \in \left[ { - {\nu_{{\rm{max}}}}/2,{\nu_{{\rm{max}}}}/{2}} \right]$. By setting ${N_{\rm{b}}} < 1/\left( {{T_{{\rm{sym}}}}{\nu _{\max }}} \right)$, we have that ${\nu _{\max }} < 1/\left( {{N_{\rm{b}}}{T_{{\rm{sym}}}}} \right)$, and  ${\nu _{u,p}} \in \left[ { - 1/(2{N_{\rm{b}}}{T_{{\rm{sym}}}}),\left. {1/(2{N_{\rm{b}}}{T_{{\rm{sym}}}})} \right]} \right.$.
	Define ${{\vartheta} _{{n}}}{\rm{ }} \buildrel \Delta \over = \frac{{{n} - {N_\upvartheta }/2}}{{{N_\upvartheta }}}$, ${{\tau} _n} \buildrel \Delta \over = \frac{n}{{{N_\uptau }\Delta f}}$, ${{\nu} _n} \buildrel \Delta \over = \frac{{\left( {n - {N_\upnu }/2} \right)}}{{{N_\upnu }{N_{\rm b}}{T_{{\rm{sym}}}}}}$, and denote the sets of the discretized grids ${\vartheta _{{n}}}$, ${\tau _{{n}}}$ and ${\nu _{{n}}}$ as  ${{\cal S}_\upvartheta } $ $\buildrel \Delta \over = $ $\left\{ {{\vartheta _0},} \right.$${\vartheta _1},$$\cdots,$$\left. {{\vartheta _{{N_\upvartheta } - 1}}} \right\}$, ${{\cal S}_\uptau } $ $\buildrel \Delta \over = $ $\left\{ {{\tau _0},} \right.$${\tau _1},$$\cdots,$$\left. {{\tau _{{N_\uptau } - 1}}} \right\}$ and ${{\cal S}_\upnu } $ $\buildrel \Delta \over = $ $\left\{ {{\nu _0},} \right.$${\nu _1},$$\cdots,$$\left. {{\nu _{{N_\upnu } - 1}}} \right\}$, where  $N_\upvartheta$, $N_\uptau$, and $N_\upnu$ are the numbers of points of the grids $\vartheta_n$, $\tau_n$, and $\nu_n$ in their ranges given above, respectively. Since all the discretized parameters ${{\vartheta _{{n}}}}$, ${{\tau _{{n}}}}$ and ${{\nu _{{n}}}}$ are acquired by sampling $\left[ { - 0.5,0.5} \right]$, $[0,1/\Delta f]$ and $\left[ { - 1/(2{N_{\rm b}}{T_{{\rm{sym}}}}),\left. {1/(2{N_{\rm b}}{T_{{\rm{sym}}}})} \right]} \right.$, respectively, we call ${{\cal S}_{{\rm{TB}}}} \buildrel \Delta \over = {{\cal S}_\upvartheta } \times {{\cal S}_\uptau } \times {{\cal S}_\upnu }$ as the range of the triple-beam (TB) domain. When ${N_\upvartheta }$, ${N_{{\uptau }}}$ and ${N_\upnu }$ are sufficiently large, we use 
	\begin{subequations}
		\begin{align}
			&{{\bar \vartheta }_{u,p}} = \mathop {\min }\limits_{\vartheta  \in {S_\upvartheta }} \left| {{\vartheta _{u,p}} - \vartheta } \right|,\label{8a}\\&{{\bar \tau }_{u,p}} = \mathop {\min }\limits_{\tau  \in {S_\uptau }} \left| {{\tau _{u,p}} - \tau } \right|,\\&{{\bar \nu}_{u,p}} = \mathop {\min }\limits_{\nu  \in {S_\upnu}} \left| {{\nu _{u,p}} - \nu } \right|,\label{8c}
		\end{align}
	\end{subequations}
	to approximate ${{\vartheta _{u,p}}}$, ${{\tau _{u,p}}}$ and ${{\nu _{u,p}}}$, respectively. With ${{\cal S}_u} \buildrel \Delta \over = {\rm{\{ }}({\bar\vartheta _{u,0}},{\bar\tau _{u,0}},{\bar\nu _{u,0}}),({\bar\vartheta _{u,1}},{\bar\tau _{u,1}},{\bar\nu _{u,1}}), \cdots ,$ $({\bar\vartheta _{u,{P_u} - 1}},{\bar\tau _{u,{P_u} - 1}},{\bar\nu _{u,{P_u} - 1}})\} $ as the parameter set of different paths, we have ${{\cal S}_u} \subseteq {{\cal S}_{{\rm{TB}}}}$. 
	
	According to the frame structure, at the time slot $T$, the indices of all the OFDM symbols in the current frame are included in ${{\cal P}_{T}}$. At the BS, we define the following steering vectors 
	\begin{align}
		&\scalebox{0.92}{$\!\!{{\bf{v}}^{\rm{s}}}\left( \vartheta  \right)\! \buildrel \Delta \over = {\left[ {1,{e^{ - \bar \jmath 2\pi \vartheta }}, \cdots ,{e^{ - \bar \jmath 2\pi (M - 1)\vartheta }}} \right]^{\rm{T}}},$}\vspace{1ex}\\
		&\scalebox{0.92}{$\!\!{{\bf{v}}^{\rm{f}}}\left( \tau  \right)\! \buildrel \Delta \over = {\left[ {{e^{ - \bar \jmath 2\pi {k_0}\Delta f\tau }}\!\!,{e^{ - \bar \jmath 2\pi ({k_0+1})\Delta f\tau }}\!\!, \cdots ,\!{e^{ - \bar \jmath 2\pi ({k_0+{{K} - 1}})\Delta f\tau }}} \right]^{\rm{T}}},$}\vspace{1ex}\\
		&\scalebox{0.92}{$\!\!{\bf{v}}_T^{\rm{t}}\left( \nu  \right){\rm{ }}\!\! \buildrel \Delta \over = {e^{\bar \jmath 2\pi {n_T}{N_{\rm b}} \nu {T_{{\rm{sym}}}}}}.{\left[ {1,{e^{\bar \jmath 2\pi \nu {T_{{\rm{sym}}}}}}\!, \cdots \!,\!{e^{\bar \jmath 2\pi \nu (N_{\rm s} - 1){T_{{\rm{sym}}}}}}}\! \right]^{\rm{T}}},$}\label{vtt}
	\end{align}
	which correspond to physical beams in the spatial, frequency and time domains, directed towards the directional cosine $\vartheta $, the delay $\tau$ and the Doppler frequency $\nu $, respectively. Therefore in the following context, we call ${{\cal S}_\upvartheta }$, ${{\cal S}_{\uptau} }$ and ${{\cal S}_\upnu }$ as the spatial beam domain, the frequency beam domain and the time beam domain, respectively, aligning closely with the aforementioned triple-beam concept. 

	With the steering vectors, at the time slot $T$, the SFT domain channel between the $u$-th UT and the BS in the current frame can be expressed as an $M \times {K} \times N_{\rm s}$ tensor, i.e, 
	\begin{align}\label{sft1}
		\scalebox{0.98}{${\mathbfcal H}_{u,T}^{{\rm{SFT}}}$}&\scalebox{0.98}{$=\displaystyle\sum\limits_{p = 0}^{{P_u} - 1} {{\alpha _{u,p}}{{\bf{v}}^{\rm{s}}}({\vartheta _{u,p}})\circ {{\bf{v}}^{\rm{f}}}\left( {{\tau _{u,p}}} \right)\circ {{\bf{v}}_T^{\rm{t}}}\left( {{\nu _{u,p}}} \right)},$}\notag\vspace{-0.2ex}\\ &\scalebox{0.98}{$\approx\displaystyle\sum\limits_{p = 0}^{{P_u} - 1} {{\alpha _{u,p}}{{\bf{v}}^{\rm{s}}}({\bar\vartheta _{u,p}})\circ {{\bf{v}}^{\rm{f}}}\left( {{\bar\tau _{u,p}}} \right)\circ {{\bf{v}}_T^{\rm{t}}}\left( {{\bar\nu _{u,p}}} \right)},$}
	\end{align}
	and the $(m,k,n)$-th element of ${\mathbfcal H}_{u,T}^{{\rm{SFT}}}$ is ${h_{u,m,k,n+{n_T}{N_{
					\rm b}}}^{\rm{SFT}}}$. 
	Furthermore, with the definition 
	\begin{equation}\label{alphaall}
		\begin{aligned}
			&\scalebox{1.1}{$ {{{\alpha _{u,n_{\upvartheta},n_{\uptau},n_{\upnu}}}}}= $}\\
			&\scalebox{0.98}{$ \begin{cases}{{\sum\limits_{p:\left( {{{\bar \vartheta }_{{u,p}}},{{\bar \tau }_{{u,p }}},{{\bar \nu }_{{u,p}}}} \right)=\left( {{\vartheta _{{n_\upvartheta }}},{\tau _{{n_\uptau }}},{\nu _{{n_\upnu }}}} \right)}\!\!\!\!\!\!\!\!\!\!\!\!\!\!\! {{\alpha _{u,p}}} }}, &\!\!\! \left(\vartheta_{n_\upvartheta}, \tau_{n_\uptau}, \nu_{n_\upnu}\right) \in \mathcal{S}_u, \\
					\!\!\quad\quad\quad\quad\quad\quad\quad\quad\quad\quad\quad\quad	\!\!\!\!0, &\!\!\! {({\vartheta _{{n_\upvartheta }}},{\tau _{{n_\uptau }}},{\nu _{{n_\upnu }}}) \notin {{\cal S}_u},}\end{cases}$}
		\end{aligned}
	\end{equation}
	where ${\bar \vartheta }_{u,p}$, ${\bar \tau }_{u,p}$, ${\bar \nu}_{u,p}$ are defined in (\ref{8a})-(\ref{8c}), ${\mathbfcal H}_{u,T}^{{\rm{SFT}}}$ in (\ref{sft1}) can be rewritten as
	\begin{align}\label{sftfunction}
		\!\!\!\!\!\!\scalebox{0.92}{$\quad {\mathbfcal H}_{u,T}^{{\rm{SFT}}}$}\scalebox{0.92}{$= \displaystyle\sum\limits_{{n_\upvartheta } = 0}^{{N_\upvartheta } - 1} \!{\sum\limits_{{n_\uptau } = 0}^{{N_\uptau } - 1}\! {\sum\limits_{{n_\upnu } = 0}^{{N_\upnu } - 1}\!\!\! {{{{\alpha _{u,n_{\upvartheta},n_{\uptau},n_{\upnu}}}}}\!{{\bf{v}}^{\rm{s}}}({\vartheta _{{n_\upvartheta }}})\!\circ {{\bf{v}}^{\rm{f}}}\left( {{\tau _{{n_\uptau }}}} \right)\circ {{\bf{v}}_T^{\rm{t}}}\left( {{\nu _{{n_\upnu }}}} \right)} } } $}.
	\end{align}
	
	By packing all the sampled steering vectors ${{{\bf{v}}^{\rm{s}}}({\vartheta _{{n_\upvartheta }}})}$, ${{{\bf{v}}^{\rm{f}}}\left( {{\tau _{{n_\uptau }}}} \right)}$ and ${{{\bf{v}}_T^{\rm{t}}}\left( {{\nu _{{n_\upnu }}}} \right)}$ into beam matrices ${{\bf{V}}^{\rm{s}}} \buildrel \Delta \over = [{{\bf{v}}^{\rm{s}}}({\vartheta _0}),{{\bf{v}}^{\rm{s}}}({\vartheta _1}), \cdots ,{{\bf{v}}^{\rm{s}}}({\vartheta _{{N_\upvartheta } - 1}})]$$\in$ $ {{\mathbb{C}}^{M \times {N_\upvartheta }}}$, ${{\bf{V}}^{\rm{f}}} \buildrel \Delta \over = [{{\bf{v}}^{\rm{f}}}({\tau _0}),{{\bf{v}}^{\rm{f}}}({\tau _1}), \cdots ,{{\bf{v}}^{\rm{f}}}({\tau _{{N_\uptau } - 1}})]$ $\in$$ {{\mathbb{C}}^{{K} \times {N_\uptau }}}$ and ${{\bf{V}}_T^{\rm{t}}} \buildrel \Delta \over = [{{\bf{v}}_T^{\rm{t}}}({\nu _0}),{{\bf{v}}_T^{\rm{t}}}({\nu _1}), \cdots ,{{\bf{v}}_T^{\rm{t}}}({\nu _{{N_\upnu } - 1}})]$ $\in $ ${{\mathbb{C}}^{N_{\rm s} \times {N_\upnu }}}$, the expression in (\ref{sftfunction}) becomes
	\begin{equation}\label{tbtosft}
		{\mathbfcal H}_{u,T}^{{\rm{SFT}}} = {{\bf{V}}_T^{\rm{t}}}{ \times _3}\left( {{{\bf{V}}^{\rm{f}}}{ \times _2}\left( {{{\bf{V}}^{\rm{s}}}{ \times _1}{\mathbfcal H}_u^{{\rm{TB}}}} \right)} \right),
	\end{equation}
	where ${\mathbfcal H}_u^{{\rm{TB}}} \in {{\mathbb{C}}^{{N_{{\rm{\upvartheta}}}} \times {N_{{\rm{\uptau}}}} \times {N_{{\rm{\upnu}}}}}}$ whose $({n_{\upvartheta}}, {n_{\uptau}}, {n_{\upnu}})$-th element is $ {{{\alpha _{u,n_{\upvartheta},n_{\uptau},n_{\upnu}}}}}$. We call ${\mathbfcal H}_u^{{\rm{TB}}}$ as the TB domain channel tensor and (\ref{tbtosft}) as the TB-based channel tensor model. With the Einstein product defined in (\ref{einstein}), such a channel model can also be expressed as
	\begin{equation}\label{eptbtosft}
		{\mathbfcal H}_{u,T}^{{\rm{SFT}}} = {{\mathbfcal V}}_T{ * _{3}}{\mathbfcal H}_u^{{\rm{TB}}},
	\end{equation}
	where \scalebox{0.94}{${{\mathbfcal V}}_T \in {{\mathbb{C}}^{M \times {K} \times {N_{\rm s}} \times {N_\upvartheta } \times {N_\uptau } \times {N_\upnu }}}$} and
	\begin{equation}\label{eptbtosft2}
		\!\!\!\!\!\!\scalebox{0.96}{$	{\left[ {{{\mathbfcal V}}}_T \right]_{m,{k},{n},{n_\upvartheta },{n_\uptau },{n_\upnu }}} = {\big[ {{{\bf{V}}^{\rm{s}}}} \big]_{m,{n_\upvartheta }}} {{\big[ {{{\bf{V}}^{\rm{f}}}} \big]_{{k},{n_\uptau }}}} {\big[ {{{\bf{V}}_T^{\rm{t}}}} \big]_{{n},{n_\upnu }}}.$}
	\end{equation}
	Since each column of ${{\bf{V}}^{\rm{s}}}$, ${{\bf{V}}^{\rm{f}}}$ and ${{\bf{V}}_T^{{\rm{t}}}}$ corresponds to a physical beam in the spatial domain, frequency domain or time domain, we call them as beam matrices, and $ {{{\mathbfcal V}}}_T$ in (\ref{eptbtosft})-(\ref{eptbtosft2}) as a triple-beam tensor.

	From the proposed frame structure in Fig. \ref{multimodel}, the set of OFDM symbols at the ${N_{\rm{p}}}$ pilot segments is denoted by ${\cal P}_{T}^{\rm{p}}$. By selecting the pilot segments from the current frame, the channel of the pilot segments can then be written as an $M\times K \times N_{\rm p}$ tensor, i.e.,
	\begin{equation}\label{eeptbtosft}
		\!\!\!\!	{\mathbfcal H}_{u,T}^{{\rm{SFT}},{\rm{p}}} = {{\bf{V}}_T^{{\rm{t}},{\rm{p}}}}{ \times _3}\left( {{\bf{V}}_{}^{\rm{f}}{ \times _2}\left( {{{\bf{V}}^{\rm{s}}}{ \times _1}{\mathbfcal H}_u^{{\rm{TB}}}} \right)} \right) = {{\mathbfcal V}_T^{\rm{p}}}{*_3}{\mathbfcal H}_u^{{\rm{TB}}},
	\end{equation}
	where the added superscript $\rm p$ stands for ``pilot'', ${\bf{V}}_T^{{\rm{t,p}}} \buildrel \Delta \over= {\left[ {{\bf{V}}_T^{\rm{t}}} \right]_{0:{N_{\rm{b}}}:{N_{\rm{s}}} - 1,:}}$$\in $ ${{\mathbb{C}}^{N_{\rm p} \times {N_\upnu }}}$, ${{\mathbfcal V}^{\rm{p}}} \in {\mathbb C}{^{M \times {K} \times {N_{\rm{p}}} \times {N_\upvartheta } \times {N_\uptau } \times {N_\upnu }}}$ and ${[{{\mathbfcal V}_T^{\rm{p}}}]_{m,k,n,{n_\upvartheta },{n_\uptau },{n_\upnu }}} = {[{{\bf{V}}^{\rm{s}}}]_{m,{n_\upvartheta }}}{[{{\bf{V}}^{\rm{f}}}]_{k,{n_\uptau }}}{[{{\bf{V}}_T^{{\rm{t}},{\rm{p}}}}]_{n,{n_\upnu }}}$. With ${{\bf{F}}_{N,g}} \buildrel \Delta \over = {{\bf{F}}_N}{{\bf{\Gamma }}_{N,g}}$, when we set ${N_\upvartheta } = {F_\upvartheta }M$,  ${N_\uptau } = {F_\uptau }{K}$, ${N_\upnu } = {F_\upnu }N_{\rm p}$, and the fine factors ${F_\upvartheta },{F_\uptau },{F_\upnu } \ge 1$, ${{\bf{V}}^{\rm{s}}}$, ${{\bf{V}}^{\rm{f}}}$ and ${{\bf{V}}_T^{{\rm{t,p}}}}$ can all be generated from the DFT matrices or partial DFT matrices, i.e., ${{\bf{V}}^{\rm{s}}} = {\left[ {{{\bf{F}}_{{N_\upvartheta },{N_\upvartheta }/2}}} \right]_{0:M - 1,:}}$, ${{\bf{V}}^{\rm{f}}} = {\left[ {{{\bf{F}}_{{N_\uptau }}}} \right]_{{\cal K},:}}$ and ${\bf{V}}_T^{{\rm{t}},{\rm{p}}} = {[{{\bf{\Gamma }}_{{N_\upnu },{n_T}}}{\bf{F}}_{{N_\upnu },{N_\upnu }/2}^*]_{{\rm{0}}:N_{\rm p} - 1,:}}$.

	Assuming the channel coefficients at different beams satisfy independent circular symmetric complex Gaussian distributions with zero mean and different variances \cite{6155694,868476}, we define the following tensor
	\begin{equation}\label{Wu}
		{{\mathbfcal W}_{{u}}} = {\mathbb{E}}\left\{ {{\mathbfcal H}_u^{{\rm{TB}}} \odot {{\left( {{\mathbfcal H}_u^{{\rm{TB}}}} \right)}^*}} \right\},
	\end{equation}
	whose $({n_{\rm{\upvartheta}}}, {n_{\rm{\uptau}}}, {n_{\rm{\upnu}}})$-th element is the power of the $({n_{\rm{\upvartheta}}}, {n_{\rm{\uptau}}}, {n_{\rm{\upnu}}})$-th beam in the TB domain. Similar to the covariance matrices of vectors, tensors have their corresponding second-order statistics. The covariance of the $3$-mode tensor ${\mathbfcal H}_{u,T}^{{\rm{SFT}},{\rm{p}}}$ can be represented as a $6$-mode tensor  \cite{schreier2010statistical}, i.e., ${\mathbfcal R}_{u,T}^{{\rm{SFT}},{\rm{p}}} \buildrel \Delta \over = {\mathbb{E}}\left\{ {{\mathbfcal H}_{u,T}^{{\rm{SFT}},{\rm{p}}} \circ {{\left( {\mathbfcal H}_{u,T}^{{\rm{SFT}},{\rm{p}}} \right)}^*}} \right\} \in {{\mathbb{C}}^{M \times {K} \times {N_{\rm{p}}} \times M \times {K} \times {N_{\rm{p}}}}}$. Meanwhile in the TB domain, the covariance tensor can also be defined as ${\mathbfcal R}_u^{{\rm{TB}}} \buildrel \Delta \over = {\mathbb{E}}\left\{ {{\mathbfcal H}_u^{{\rm{TB}}} \circ {{\left( {{\mathbfcal H}_u^{{\rm{TB}}}} \right)}^*}} \right\} \in {{\mathbb{C}}^{{N_\upvartheta } \times {N_\uptau } \times {N_\upnu } \times {N_\upvartheta } \times {N_\uptau } \times {N_\upnu }}}$. Note that ${\mathbfcal R}_{u}^{{\rm{TB}}}$ is pseudo-diagonal, and the pseudo-diagonal elements  satisfy ${\left[ {{\mathbfcal R}_u^{{\rm{TB}}}} \right]_{{n_1},{n_2},{n_3},{n_1},{n_2},{n_3}}} = {\left[ {{\mathbfcal W}_u^{}} \right]_{{n_1},{n_2},{n_3}}}$. With the relationship in (\ref{eeptbtosft}), ${\mathbfcal R}_{u,T}^{{\rm{SFT}},{\rm{p}}}$ can be calculated from ${\mathbfcal R}_u^{{\rm{TB}}}$ by
	\begin{equation}\label{diag}
		{\mathbfcal R}_{u,T}^{{\rm{SFT}},{\rm{p}}} = {{\mathbfcal V}}_T^{{\rm{p}}}{ * _3}{\mathbfcal R}_u^{{\rm{TB}}} {*_3} \left( {{{{\mathbfcal V}}_T^{\rm{p}}}} \right)_3^{\rm{H}}.
	\end{equation}
	
	Note that the three dimensions of the TB domain channel tensor ${\mathbfcal H}_u^{\rm{TB}}$ correspond to the spatial beam, frequency beam and time beam, respectively.  From (\ref{sft1})-(\ref{alphaall}), ${\mathbfcal H}_u^{\rm{TB}}$ is supported on set ${\cal S}_u$, i.e., its components are only non-zero on ${\cal S}_u$. Based on the OFDM design principle, the channel delay spread width is no larger than the CP length and ${N_{\rm{g}}}{T_{\rm{s}}} \le {\tau _{{F_\uptau }{N_{\rm{f}}}}}$, where ${N_{\rm{f}}} \buildrel \Delta \over = \left\lceil {{N_{\rm{g}}}{K}/{N_{\rm{c}}}} \right\rceil $. Thus the non-zero supports of ${\mathbfcal H}_u^{\rm{TB}}$ are located within $\left[ {0,{F_\uptau }{N_{\rm{f}}}} \right)$ along the frequency beam dimension. Similarly, since the Doppler spread width satisfies ${\nu _{\max }}/2 \le {\nu _{{F_\upnu }{N_{\rm{d}}}}}$, where  ${N_{\rm{d}}} \buildrel \Delta \over = \left\lceil {{\nu _{\max }}{T_{{\rm{sym}}}}N_{\rm s}} \right\rceil $, the non-zero supports of ${\mathbfcal H}_u^{\rm{TB}}$ are located within $\left[ {({N_\upnu } - {N_{\rm{d}}}{F_\upnu })/2:({N_\upnu } + {N_{\rm{d}}}{F_\upnu })/2} \right)$ along the time beam dimension. Therefore, ${\cal S}_u$ should be a small subset of ${\cal S}_{\rm {TB}}$, and ${\mathbfcal H}_u^{\rm{TB}}$ is a sparse tensor.
	From (\ref{Wu}), ${\mathbfcal W}_u$ has the same sparsity as ${\mathbfcal H}_u^{{\rm{TB}}}$. Besides, as a kind of statistical CSI, ${\mathbfcal W}_{u}$ varies relatively slowly as time evolves \cite{9967937}. Hence, abundant  resources can be adopted for acquiring ${\mathbfcal W}_u$, and we assume that ${\mathbfcal W}_u$ of all the UTs are available at the BS in the rest of this paper.

	\section{Time-frequency Phase-shifted Pilot} \label{section3}
	
	
	In this section, we propose our TFPSP design and the corresponding pilot scheduling method based on the TB-based channel tensor model. The proposed pilot design differs significantly from conventional frequency-domain pilots. It is closely related to the TB-based channel tensor model discussed earlier, aiming to address the shortcomings of existing pilot design methods in terms of pilot overhead and pilot interference.

	\subsection{TFPSP}

	Before proceeding with our pilot design, we first discuss the widely-used frequency domain phase-shifted pilots for massive MIMO-OFDM systems. With ${\bf{x}}_u^{\rm{f}} \in {{\mathbb{C}}^{{K} \times 1}}$ as the pilot sequence transmitted from the $u$-th UT and ${[{\bf{x}}_u^{\rm{f}}]_{k}}$ as the signal transmitted on the $(k+k_0)$-th subcarrier, ${\bf{x}}_u^{\rm{f}}$ can be designed as
	\begin{equation}\label{Xuf}
		{\bf{x}}_u^{\rm{f}} = {\bf{f}}_{{\phi _u}}^{\rm{f}} \odot {{\bf{x}}^{\rm{f}}},
	\end{equation}
	where ${\phi _u} \in \{ 0,1, \cdots ,{K} - 1\} $ represents the phase shift of the $u$-th UT, and ${\bf{f}}_{{\phi _u}}^{\rm{f}} = \left[ {{e^{ - {\bar \jmath }2\pi {k_0}\frac{{{\phi _u}}}{K}}},{e^{ - {\bar \jmath }2\pi \left( {{k_0}{\rm{ + 1}}} \right)\frac{{{\phi _u}}}{K}}},} \right.$$\cdots$ $\left. {,{e^{ - {\bar \jmath }2\pi \left( {{k_0} + K - 1} \right)\frac{{{\phi _u}}}{K}}}} \right]^{\rm T}$, which can be generated from the ${K} \times {K}$ DFT matrix, i.e., ${\bf{f}}_{{\phi _ u}}^{\mathop{\rm f}\nolimits} = {e^{ - \bar \jmath 2\pi {k_0}\frac{{{\phi_ u}}}{{K}}}}{{\bf{f}}_{{K},{\phi_u}}}$. ${{\bf{x}}^{\rm{f}}}$ is a basic pilot sequence shared by all UTs, whose components are all unit modulus, and a Zadoff-Chu (ZC) sequence is usually used to generate ${{\bf{x}}^{\rm{f}}}$ \cite{Chu1972,dahlman20205g}. Considering that the delay spread should be no larger than ${\tau _{{F_\uptau }{N_{\rm{f}}}}}$ for the massive MIMO-OFDM system, then to minimize the mean square error (MSE) of channel estimation, one sufficient condition is the phase-shifted orthogonality \cite{1200150}, i.e., for $\forall u,u' \in {{\cal U} }$, $u \ne u'$,
	\begin{equation}\label{psopcondition}
		{\rm{tr}}\left\{ {{\bf{X}}_u^{\rm{f}}{\rm{Diag}}\left\{ {{\bf{f}}_\phi ^{\rm{f}}} \right\}{{\left( {{\bf{X}}_{u'}^{\rm{f}}} \right)}^{\rm{H}}}} \right\} = 0,\forall \left| \phi  \right| \le {N_{\rm{f}}},
	\end{equation}
	where ${\bf{X}}_u^{\rm{f}} = {\rm{Diag}}\left\{ {{\bf{x}}_u^{\rm{f}}} \right\}$. This condition can be satisfied if and only if $\forall u,u' \in {{\cal U} }$, $u \ne u'$, $\left| { {{\phi _{u'}} - {\phi _u}}} \right| \ge {N_{\rm{f}}}$ holds, and one choice is ${\phi _u} = u{N_{\rm{f}}}$. Thus, for such phase-shifted orthogonal pilots (PSOPs), there are only $\left\lfloor {K}/{N_{\rm f}} \right\rfloor $ available phase shifts that can be scheduled for different UTs. Meanwhile, with 
	${\bf{X}}_{u,u'}^{\rm{f}} \buildrel \Delta \over = {\left( {{\bf{X}}_u^{\rm{f}}} \right)^{\rm{H}}}{\bf{X}}_{u'}^{\rm{f}}$, we have ${\bf{X}}_{u,u'}^{\rm{f}} = {\rm{Diag}}\left\{ {{\bf{f}}_{{\phi _{u'}} - {\phi _u}}^{\rm{f}}} \right\}$.

	\begin{figure}[t]
		\centering
		\subfloat[FPSPs transmitted in the current frame at the time slot $T$. ]{\includegraphics[width=242pt]{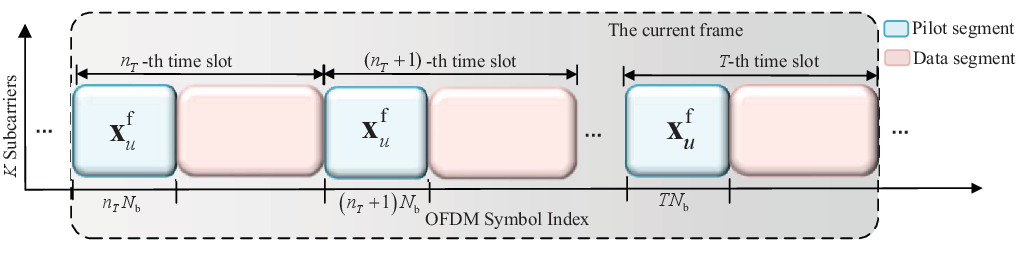}\label{pilot1}}\\
		\subfloat[TFPSPs transmitted in the current frame at the time slot $T$.]{\includegraphics[width=250pt]{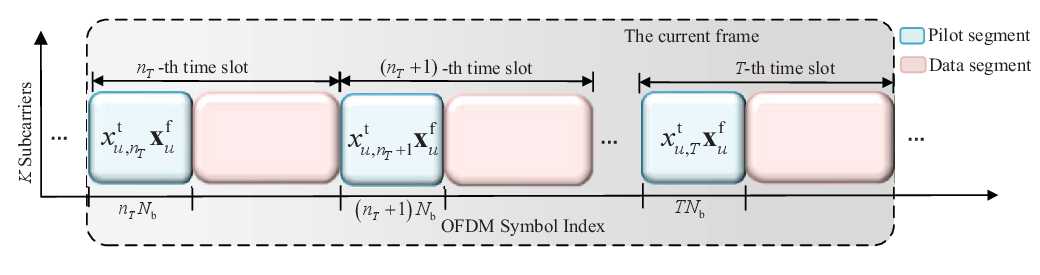}\label{pilot2}}\\
		\caption{Pilots transmitted in the current frame at the time slot $T$. }\label{figure4}
	\end{figure}

	PSOPs have already been adopted in LTE and 5G NR \cite{dahlman20205g} due to their good estimation performance. However, the above phase restriction for PSOPs limits its application to accommodate a large number of UTs. Although increasing the number of pilot segments  can generate more available pilots, it leads to an increase in pilot overhead and a decrease in spectrum efficiency \cite{Marzetta2010non}.
	To overcome these problems, adjustable phase-shifted pilots (APSPs) in \cite{7332961} can be used instead of PSOPs. In this case, the phase shifts are adjustable, and the orthogonality in (\ref{psopcondition}) is not necessary, i.e.,  $\forall u,u' \in {{\cal U} }$, $u \ne u'$, $\left| { {{\phi _{u'}} - {\phi _u}} } \right|$ need not to be larger than $N_{\rm f}$. Hence, the application of APSPs provide more available pilot resources without increasing the pilot overhead.  \cite{7332961} also provided the condition, under which APSPs can reach the same performance as PSOPs by scheduling the phase shifts properly. In this case, some UTs can even share the same phase shift, verifying the feasibility of pilot reuse \cite{you2015pilot, 9042356}.

	
	As shown above, both PSOPs and APSPs are transmitted along the frequency dimension conventionally, which we refer to as frequency phase-shifted pilots (FPSPs). Applying FPSPs to the frame structure in Fig. \ref{multimodel}, at the time slot $T$, all the OFDM symbols in the pilot segments of the current frame transmit the same ${\bf{x}}_u^{\rm{f}}$ over the valid subcarriers, as shown in Fig. \ref{figure4} (a). With FPSPs, the received signal at the BS can be expressed as an $M\times K \times N_{\rm p}$ tensor, i.e.,
	\begin{align}\label{Yuf}
		{{\mathbfcal Y}_T^{\rm{p}}}{\rm{ }} = \sqrt {{\sigma _{\rm{p}}}} \sum\limits_{u = 0}^{U - 1} {{\bf{X}}_{u}^{\rm{f}}{ \times _2}{\mathbfcal H}_{u,T}^{{\rm{SFT}},{\rm{p}}}}  + {\mathbfcal Z},
	\end{align}
	where ${[{{\mathbfcal Y}_T^{\rm{p}}}]_{:,k,n}}$ is the received signal at the BS of the $k$-th subcarrier on the $n$-th pilot segment of the frame, ${{\sigma _{\rm{p}}}}$ is the transmit signal power and ${\mathbfcal{Z}}$ is the additive white Gaussian noise (AWGN) tensor with each element identically and independently distributed as ${\cal C}{\cal N}\left( {{0}},\sigma _z \right)$.
	
	With the specific channel characteristics in the delay domain, where the delay spread is small relative to $1/\Delta f$, extensive research has been conducted on the channel estimation with FPSPs. Interestingly, similar channel characteristics can also be exploited in the Doppler domain, where the Doppler spread is small relative to $1/\left( {{N_{\rm{b}}}{T_{{\rm{sym}}}}} \right)$ and typically sparsity exists \cite{Shen, Farhang2018otfs}.  Accordingly, phase shift pilots can also be transmitted in the time domain. In this case, as illustrated in Fig. \ref{figure4} (b), ${{\bf{x}}_u^{\rm{f}}}$ transmitted in the pilot segment of the $n$-th slot is multiplied by $x_{u,n}^{\rm{t}}$, which may be different for different $n$. Then, along the time dimension, all the $x_{u,n}^{\rm{t}}$ for $n = 0,1,2,\cdots$ constitute the time domain pilot sequence. Similar to the design of ${\bf{x}}_{u}^{\rm{f}}$ in (\ref{Xuf}), $x_{u,n}^{\rm{t}}$ can also be generated by a phase shift sequence and a ZC sequence, i.e., 
	\begin{equation}
		x_{u,n}^{\rm{t}} = {e^{ - \bar \jmath 2\pi n\frac{{{\varphi _u}}}{{{N_{\rm{p}}}}}}} \cdot {e^{\bar \jmath 2\pi \gamma \frac{{n\left( {n + {{\left\langle {{N_{\rm{p}}}} \right\rangle }_2}} \right)}}{{{N_{\rm{p}}}}}}}, n=0,1,2,\cdots,
	\end{equation} 
	where $\gamma$ is an integer that is coprime to $N_{\rm p}$ \cite{Chu1972}, and the phase shift ${\varphi _u} \in {\rm{ }}\left\{ {0,1, \cdots ,} \right.$ $\left. {{N_{\rm{p}}} - 1} \right\}$. Under this setting, we have $x_{u,n}^{\rm{t}} = x_{u,n + {N_{\rm{p}}}}^{\rm{t}}$, indicating that $x_{u,n}^{\rm{t}}$ is a periodic sequence with period ${N_{\rm{p}}}$. At the time slot $T$, the pilot signal of the current frame is ${\bf{x}}_{u,T}^{\rm{t}} \buildrel \Delta \over = {[{{x}}_{u,{n_T}}^{\rm{t}},{{x}}_{u,{n_T} + 1}^{\rm{t}}, \cdots ,{{x}}_{u,T}^{\rm{t}}]^{\rm{T}}}$. Due to the periodicity of $x_{u,n}^{\rm{t}}$, we can express ${\bf{x}}_{u,T}^{\rm{t}}$ as
	\begin{equation}\label{Xut}
		{\bf{x}}_{u,T}^{\rm{t}} = {{\bf{\Gamma }}_{{N_{\rm{p}}},{n_T}}}\left( {{\bf{f}}_{{\varphi _u}}^{\rm{t}} \odot {{\bf{x}}^{\rm{t}}}} \right),
	\end{equation}
	where ${\bf{f}}_{{\varphi _u}}^{\rm{t}} = {\bf{f}}_{{N_{\rm{p}}},{{ {{\varphi _u}} }}}$ and ${{\bf{x}}^{\rm{t}}}$ is composed of the ${N_{\rm p}}$-point ZC sequence with ${[{{\bf{x}}^{\rm{t}}}]_n} = {e^{{{\bar \jmath 2\pi \gamma n\left( {n + {{\left\langle {{N_{\rm{p}}}} \right\rangle }_2}} \right)} \mathord{\left/
					{\vphantom {{\bar \jmath 2\pi \gamma n\left( {n + {{\left\langle {{N_{\rm{p}}}} \right\rangle }_2}} \right)} {{N_{\rm{p}}}}}} \right.
					\kern-\nulldelimiterspace} {{N_{\rm{p}}}}}}}$.   
		Defining ${\bf{X}}_{u,T}^{\rm{t}} = {\rm{Diag}}\left\{ {{\bf{x}}_{u,T}^{\rm{t}}} \right\}$, ${\bf{X}}_{u,u',T}^{\rm{t}} \buildrel \Delta \over = {\left( {{\bf{X}}_{u,T}^{\rm{t}}} \right)^{\rm{H}}}{\bf{X}}_{u',T}^{\rm{t}}$, for $\forall u,u' \in {{\cal U} }$, $u \ne u'$, we have that ${\bf{X}}_{u,u',T}^{\rm{t}} = {\rm{Diag}}\left\{ {{{\bf{\Gamma }}_{{N_{\rm{p}}},{n_T}}}{\bf{f}}_{{\varphi _{u'}} - {\varphi _u}}^{\rm{t}}} \right\}$. With the above pilot design, the signals transmitted on different pilot segments may vary, differing from FPSPs, and we refer to such a design as TFPSP.

		At the time slot $T$, by transmitting TFPSPs as Fig. \ref{figure4} (b) shows, the received signal at the BS can be expressed in tensor as
		\begin{align}\label{Xu}
			{{\mathbfcal Y}_T^{\rm{p}}} &= \sqrt {{\sigma _{\rm{p}}}} \sum\limits_{u = 0}^{U - 1} {{\bf{X}}_{u,T}^{\rm{t}}{ \times _3}\left( {{\bf{X}}_u^{\rm{f}}{ \times _2}{\mathbfcal H}_{u,T}^{{\rm{SFT}},{\rm{p}}}} \right)}  + {\mathbfcal Z}\notag\\
			&= \sqrt {{\sigma _{\rm{p}}}} \sum\limits_{u = 0}^{U - 1} {{\mathbfcal X}_{u,T}^{}{ * _3}{\mathbfcal H}_{u,T}^{{\rm{SFT}},{\rm{p}}}}  + {\mathbfcal Z},
		\end{align}
		where the TFPSP signal of the $u$-th UT is the pseudo-diagonal tensor ${\mathbfcal X}_{u,T}^{} \in {{\mathbb{C}}^{M \times {K} \times {N_{\rm{p}}} \times M \times {K} \times {N_{\rm{p}}}}}$, satisfying ${[{\mathbfcal X}_{u,T}^{}]_{m,k,{n},m,k,{n}}} = {\left[ {{\bf{X}}_u^{\rm{f}}} \right]_{k,k}}{\left[ {{\bf{X}}_{u,T}^{\rm{t}}} \right]_{n,n}}$. Compared with (\ref{Yuf}), transmitting FPSPs is equivalent to setting ${\bf{X}}_{u,T}^{\rm{t}} = {{\bf{I}}_{{N_{\rm{p}}}}}$ in (\ref{Xu}), or ${\varphi _u} = 0$ and ${\bf x}_{u,T}^{\rm t}$ in (\ref{Xut}) is the all $1$ vector $\bf 1$ for all $u \in {\cal U}$.
		
		
		
		\subsection{Optimal Condition on TFPSPs}
		
		Both ${\phi _u}$ and ${\varphi _u}$ can be scheduled in TFPSPs, and this may have an impact on channel estimation performance as we shall see below. With the relationship in (\ref{eeptbtosft}), (\ref{Xu}) can be further expressed with the TB domain channel tensor ${{\mathbfcal H}_u^{{\rm{TB}}}}$ as
		\begin{align}\label{rec1}
			{{\mathbfcal Y}_T^{\rm{p}}} = & \sqrt {{\sigma _{\rm{p}}}} {{\mathbfcal X}_{u,T}} { * _3} {\mathbfcal V}_T^{\rm{p}}{ * _3}{\mathbfcal H}_u^{{\rm{TB}}} \notag\\&+ \sqrt {{\sigma _{\rm{p}}}} \sum\limits_{u' \ne u} {{{\mathbfcal X}_{u',T}} { * _3} {\mathbfcal V}_T^{\rm{p}}{ * _3}{\mathbfcal H}_{u'}^{{\rm{TB}}}}  + {\mathbfcal Z}.
		\end{align}
		For any two different UTs $u,u' \in {\cal U}$, we define 
		\begin{align}\label{jiaocha0}
			\scalebox{0.96}{${\mathbfcal R}_{u,u',T}^{{\rm{SFT}},{\rm{p}}} \buildrel \Delta \over = \left( {{\mathbfcal X}_{u,T}^{}} \right)_3^{\rm{H}}{ * _3}{{\mathbfcal X}_{u',T}}{ * _3}{\mathbfcal R}_{u',T}^{{\rm{SFT}},{\rm{p}}}{ * _3}\left( {{\mathbfcal X}_{u',T}^{}} \right)_3^{\rm{H}}{ * _3}{\mathbfcal X}_{u,T}^{}.$}
		\end{align}
		Then by adopting the criterion of tensor-based MMSE \cite{pandey2021multi}, 
		for each UT, the estimate of ${\mathbfcal H}_u^{{\rm{TB}}}$ is given as
		\begin{equation}\label{estimateHtb}
			\scalebox{1}{$\hat {\mathbfcal H}_u^{{\rm{TB}}} = {\mathbfcal R}_u^{{\rm{TB}}}{ * _3}\left( {{{\mathbfcal V}_T^{\rm{p}}}} \right)_3^{\rm{H}}{ * _3}{\mathbfcal C}_{{\rm{u,all}}}^{ - 1}{ * _3}\left( {{{\mathbfcal X}_{u,T}}} \right)_3^{\rm{H}}{ * _3}\frac{{{{\mathbfcal Y}_T^{\rm{p}}}}}{{\sqrt {{\sigma _{\rm{p}}}} }}$},
		\end{equation}
		where ${{\mathbfcal{C}}_{u,{\rm{all}}}} \in {{\mathbb{C}}^{M \times {K} \times {N_{\rm{p}}} \times M \times {K} \times {N_{\rm{p}}}}}$ is square and 
		\begin{equation}\label{Cu}
			{\mathbfcal{C}}_{u,{\rm{all}}}^{} \buildrel \Delta \over = {\mathbfcal{R}}_{u,T}^{{\rm{SFT,p}}} + \sum\limits_{u' \ne u} {{{\mathbfcal R}_{u,u',T}^{{\rm{SFT}},{\rm{p}}}}}  + \frac{{{\sigma _z}}}{{{\sigma _{\rm{p}}}}}{{\mathbfcal{I}}_{M,{K},{N_{\rm{p}}}}}.
		\end{equation}

		We conduct an analysis of ${\mathbfcal R}_{u,u',T}^{{\rm{SFT}},{\rm{p}}}$ in (\ref{jiaocha0}), as it is highly related to the pilot signals ${\mathbfcal X}_{u,T}$ and ${\mathbfcal X}_{u',T}$. By utilizing the properties of tensor products described in  \cite{opentensor}, ${\mathbfcal R}_{u,u',T}^{{\rm{SFT}},{\rm{p}}}$ can be further expressed as
		\begin{align}\label{jiaocha1}
			{\mathbfcal R}_{u,u',T}^{{\rm{SFT}},{\rm{p}}} =  {\mathbb{E}}\Big\{{{\mathbfcal H}_{u,u',T}^{{\rm{SFT}},{\rm{p}}} \circ {{\left( {{\mathbfcal H}_{u,u',T}^{{\rm{SFT}},{\rm{p}}}} \right)}^*}} \Big\},
		\end{align}
		where 
		\begin{equation}\label{jiaocha01}
			\scalebox{0.98}{${\mathbfcal H}_{u,u',T}^{{\rm{SFT}},{\rm{p}}}{\rm{ }} \buildrel \Delta \over = {\bf{X}}_{u,u',T}^{\rm{t}}{ \times _3}\left( {{\bf{X}}_{u,u'}^{\rm{f}}{ \times _2}{\mathbfcal H}_{u',T}^{{\rm{SFT}},{\rm{p}}}} \right).$}
		\end{equation}
		In the design of TFPSPs, both ${{\bf{X}}_{u,u'}^{\rm{f}}}$ and ${{\bf{X}}_{u,u',T}^{\rm{t}}}$ in (\ref{jiaocha01}) are diagonal, and their diagonal elements ${{\bf{f}}_{{\phi _{u'}} - {\phi _u}}^{\rm{f}}}$ and ${{{\bf{\Gamma }}_{{N_{\rm{p}}},{n_T}}}{\bf{f}}_{{\varphi _{u'}} - {\varphi _u}}^{\rm{t}}}$ can be generated from the columns of DFT matrices. Meanwhile, since the beam matrices ${{\bf{V}}^{\rm{f}}}$ and ${\bf{V}}_T^{{\rm{t,p}}}$ can be derived from DFT or partial DFT matrices, ${{\bf{f}}_{{\phi _{u'}} - {\phi _u}}^{\rm{f}}}$ and ${{{\bf{\Gamma }}_{{N_{\rm{p}}},{n_T}}}{\bf{f}}_{{\varphi _{u'}} - {\varphi _u}}^{\rm{t}}}$ can also be represented with the columns of ${{\bf{V}}^{\rm{f}}}$ and ${\bf{V}}_T^{{\rm{t,p}}}$, respectively. Hence, we can establish the following relationships ${\bf{X}}_{u,u'}^{\rm{f}}{{\bf{V}}^{\rm{f}}} = {{\bf{V}}^{\rm{f}}}{{\bf{\Gamma }}_{{N_\uptau },\left( {{\phi _{u}} - {\phi _{u'}}} \right){F_\uptau }}}$ and $ {{\bf{X}}_{u,u',T}^{\rm{t}}{\bf{V}}_T^{{\rm{t,p}}}}$$=$$ {{\bf{V}}_T^{{\rm{t,p}}}{{\bf{\Gamma }}_{{N_\upnu },\left( {{\varphi _{u'}} - {\varphi _u}} \right){F_\upnu }}}} $. In this case, we introduce the following definition 
		\begin{equation}\label{Hab}
			{\mathbfcal H}_{u,a,b}^{{\rm{TB}}}  \buildrel \Delta \over =  {{\bf{\Gamma }}_{{N_\upnu },b}}{ \times _3}\left( {{{\bf{\Gamma }}_{{N_\uptau },a}}{ \times _2}{\mathbfcal H}_u^{{\rm{TB}}}} \right),
		\end{equation}
		where all the elements in ${\mathbfcal H}_u^{{\rm{TB}}}$ undergo the positional changes. Analogous to the cyclic shifts of sequences, we refer to the above transformation (\ref{Hab}) as the tensor cyclic shift, where $a$ and $b$ correspond to the cyclic shift lengths along its second and third dimensions,  respectively.
	With (\ref{Hab}), the term ${{\mathbfcal H}_{u,u'}^{{\rm{SFT}},{\rm{p}}}}$ can then be written by
	\begin{align}\label{jiaocha2}
		\!\!\!\!\scalebox{0.95}{${\mathbfcal H}_{u,u',T}^{{\rm{SFT}},{\rm{p}}}$} &\scalebox{0.95}{$= \left( {{\bf{X}}_{u,u',T}^{\rm{t}}{{\bf{V}}_T^{{\rm{t}},{\rm{p}}}}} \right){ \times _3}\left( {\left( {{\bf{X}}_{u,u'}^{\rm{f}}{{\bf{V}}^{\rm{f}}}} \right){ \times _2}\left( {{{\bf{V}}^{\rm{s}}}{ \times _1}{\mathbfcal H}_{u'}^{{\rm{TB}}}} \right)} \right)$}\notag\\
		& \scalebox{0.93}{$= \left( {{{\bf{V}}_T^{{\rm{t}},{\rm{p}}}}{{\bf{\Gamma }}_{{N_\uptau },\left( {{\varphi _{u'}} - {\varphi _u}} \right){F_\uptau }}}} \right)$}\notag\\&\scalebox{0.95}{$\quad\quad{ \times _3}\left( {\left( {{{\bf{V}}^{\rm{f}}}{{\bf{\Gamma }}_{{N_\uptau },\left( {{\phi _u} - {\phi _{u'}}} \right){F_\uptau }}}} \right){ \times _2}\left( {{{\bf{V}}^{\rm{s}}}{ \times _1}{\mathbfcal H}_{u'}^{{\rm{TB}}}} \right)} \right)$}\notag\\
		&\scalebox{0.93}{$= \!{{\bf{V}}_T^{{\rm{t}},{\rm{p}}}}{ \times _3}\left( \!{{{\bf{V}}^{\rm{f}}}{ \times _2}\left(\! {{{\bf{V}}^{\rm{s}}}{ \times _1}{\mathbfcal H}_{u',{L_{{\phi _{u'}}}} - {L_{{\phi _u}}},{L_{{\varphi _{u'}}}} - {L_{{\varphi _u}}}}^{{\rm{TB}}}} \right)}\! \right)$},
	\end{align}
	where ${L_{{\phi _u}}} = -{\phi _u}{F_\uptau }$ and ${L_{{\varphi _u}}} = {\varphi _u}{F_\upnu }$. 
	From (\ref{jiaocha1}), since ${\mathbfcal R}_{u,u',T}^{{\rm{SFT}},{\rm{p}}}$ can be regarded as the covariance of ${\mathbfcal H}_{u,u',T}^{{\rm{SFT}},{\rm{p}}}$, defining the following tensor	
	\begin{align}\label{Wab}
		\scalebox{0.96}{${{\mathbfcal W}_{u,a,b}}\!\!= {\mathbb{E}}\!\left\{ {{\mathbfcal H}_{u,a,b}^{{\rm{TB}}} \odot {{\left( {{\mathbfcal H}_{u,a,b}^{{\rm{TB}}}} \right)}^*}} \right\}\!\!= {{\bf{\Gamma }}_{{N_\upnu },b}}{ \times _3}\left( {{{\bf{\Gamma }}_{{N_\uptau },a}}{ \times _2}{{\mathbfcal W}_{u}}} \right),$}
	\end{align}
	${\mathbfcal R}_{u,u',T}^{{\rm{SFT}},{\rm{p}}}$ can then be expressed as
	\begin{equation}\label{jiaochalast}
		\scalebox{0.96}{${\mathbfcal{R}}_{u,u',T}^{{\rm{SFT}},{\rm{p}}} = {\mathbfcal V}_T^{\rm{p}}{ * _3}{\mathbfcal{R}}_{u,u'}^{{\rm{TB}}}{ * _3}{\left( {{\mathbfcal V}_T^{\rm{p}}} \right)^{\rm{H}}_3},$}
	\end{equation}
	where \scalebox{0.95}{${\mathbfcal R}_{u,u'}^{{\rm{TB}}}\in{{\mathbb{C}}^{{N_\upvartheta } \times {N_\uptau } \times {N_\upnu } \times {N_\upvartheta } \times {N_\uptau } \times {N_\upnu }}}$} is pseudo-diagonal, and \scalebox{0.95}{${\left[ {{\mathbfcal R}_{u,u'}^{\rm{TB}}} \right]_{{n_1},{n_2},{n_3},{n_1},{n_2},{n_3}}}\!\! = \!\!{\left[ {{\mathbfcal W}_{u',{L_{{\phi _{u'}}}} - {L_{{\phi _u}}},{L_{{\varphi _{u'}}}} - {L_{{\varphi _u}}}}^{}} \right]_{{n_1},{n_2},{n_3}}}$}.
	
	The expressions in (\ref{jiaocha2}) and (\ref{jiaochalast}) indicate that,  transmitting TFPSP signals has an affect on the TB domain channels, which is equivalent to the tensor cyclic shift, and the cyclic shift lengths are determined by the phase shifts of TFPSPs. Then, to facilitate the following expressions, we call ${\mathbfcal H}_{u,{L_{{\phi _u}}},{L_{{\varphi _u}}}}^{{\rm{TB}}}$ as the equivalent TB domain channel tensor of the $u$-th UT, and ${{\mathbfcal W}_{u,{L_{{\phi _u}}},{L_{{\varphi _u}}}}}$ as the equivalent TB domain power distribution. 

	With the relationship between ${\mathbfcal H}_{u,T}^{{\rm{SFT}},{\rm{p}}}$ and ${\mathbfcal H}_u^{{\rm{TB}}}$, after obtaining $\hat {\mathbfcal H}_u^{{\rm{TB}}}$, the MMSE estimate of ${\mathbfcal H}_{u,T}^{{\rm{SFT}},{\rm{p}}}$ is 
	\begin{equation}
		\!\!\!\!\!\scalebox{0.98}{$\hat {\mathbfcal H}_{u,T}^{{\rm{SFT,p}}} \!\!= {{\mathbfcal V}_T^{\rm{p}}}{ * _3}\hat {\mathbfcal H}_u^{{\rm{TB}}} \!\!= {\mathbfcal R}_{u,T}^{{\rm{SFT,p}}}{ * _3}{\mathbfcal C}_{{\rm{u,all}}}^{ - 1}{ * _3}\left( {{{\mathbfcal X}_{u,T}}} \right)_3^{\rm{H}}{ * _3}\frac{{{{\mathbfcal Y}_T^{\rm{p}}}}}{{\sqrt {{\sigma _{\rm{p}}}} }},$}
	\end{equation}
	and the MSE between ${\mathbfcal H}_{u,T}^{{\rm{SFT}},{\rm{p}}}$ and $\hat {\mathbfcal H}_{u,T}^{{\rm{SFT}},{\rm{p}}}$ for all the $u \in {\cal U}$ is then given by 
	\begin{equation}
		\label{NMSE2}
		\scalebox{0.98}{$\!\!\!\!\varepsilon _{{\rm{MSE}}}^{} = \sum\limits_{u = 0}^{U - 1} {{\rm{tr}}\left\{ {{\mathbfcal R}_{u,T}^{{\rm{SFT}},{\rm{p}}} - {\mathbfcal R}_{u,T}^{{\rm{SFT}},{\rm{p}}}{ * _3}{\mathbfcal C}_{u,{\rm{all}}}^{ - 1}{ * _3}{\mathbfcal R}_{u,T}^{{\rm{SFT}},{\rm{p}}}} \right\}}.$}
	\end{equation}
	It can be observed from (\ref{NMSE2}) that, the value of $\varepsilon _{{\rm{MSE}}}$ depends on the statistical CSI of each UT, as well as the interference among all the UTs in $\cal U$, which is contained in the term ${\mathbfcal{C}}_{u,{\rm{all}}}$. In order to eliminate the inter-UT interference and reduce the MSE, we have the following theorem, giving the minimum value of MSE and a condition of reaching the value.
	\begin{figure*}[!t]

		\begin{align}\label{NMSEMIN}
			\varepsilon _{{\rm{MSE}}}^{} \ge \varepsilon _{{\rm{MSE}},{\rm{min}}}^{} = \sum\limits_{u = 0}^{U - 1} {{\rm{tr}}\left\{ {{\mathbfcal{R}}_{u,T}^{{\rm{SFT}},{\rm{p}}} - {\mathbfcal{R}}_{u,T}^{{\rm{SFT}},{\rm{p}}}{ * _3}{{\left( {{\mathbfcal{R}}_{u,T}^{{\rm{SFT}},{\rm{p}}} + \frac{{{\sigma _z}}}{{\sigma _{\rm{p}}}}{{\mathbfcal{I}}_{M,{K},{N_{\rm{p}}}}}} \right)}^{ - 1}}{ * _3}{\mathbfcal{R}}_{u,T}^{{\rm{SFT}},{\rm{p}}}} \right\}}.
		\end{align}
		\hrulefill
	\end{figure*}

	\begin{theorem}\label{theorem1}
		When ${M,{K},{N_{\rm{p}}} \to \infty }$, $\varepsilon _{{\rm{MSE}}}^{}$ is lower bounded by (\ref{NMSEMIN}). This lower bound can be achieved if $\forall u,u' \in {{\cal U} }$, $u \ne u'$, 
		\begin{equation}\label{condition}
			{{\mathbfcal W}_{u,{L_{{\phi _u}}},{L_{{\varphi _u}}}}} \odot {{\mathbfcal W}_{u',{L_{{\phi _{u'}}}},{L_{{\varphi _{u'}}}}}} = {\bf{0}}.
		\end{equation}
	\end{theorem}
	\emph{Proof}: See Appendix \ref{app1}.
	
	The condition in (\ref{condition}) implys that the equivalent channel power distributions of different UTs are non-overlapping in the TB domain. Compared with ${\varepsilon _{{\rm{MSE}}}}$ in (\ref{NMSE2}), the inter-UT interference is eliminated in ${\varepsilon _{{\rm{MSE,min}}}}$.  Hence Theorem \ref{theorem1} suggests that non-overlapping equivalent channel power distributions in the TB domain lead to the elimination of inter-UT interference and the minimization of ${\varepsilon _{{\rm{MSE}}}}$. Since ${{\mathbfcal W}_{u,{L_{{\phi _u}}},{L_{{\varphi _u}}}}}$ for each $u \in {\cal U}$ is determined by the phase shifts $({\phi _u},{\varphi _u})$ of TFPSPs, the condition (\ref{condition}) can be achieved by selecting proper TFPSPs for all $u \in {\cal U}$. Actually, the existing FPSP scheduling methods follow similar principles. For example, since the delay spread is no larger than ${\tau _{{F_\uptau }{N_{\rm{f}}}}}$, PSOPs allow the seperation of equivalent channels in the delay (frequency beam) domain by setting $\left| {{\phi_{u'}} - {\phi_u}} \right| \ge {N_{\rm{f}}}$, i.e., $\left| {{L_{{\phi _{u'}}}} - {L_{{\phi _u}}}} \right| \ge {N_{\rm{f}}}{F_{\uptau}}$  for $\forall u,u' \in {{\cal U} }$, $u \ne u'$. APSPs exploit the channel sparsity in the angular-delay (spatial-frequency beam) and achieve such seperation with adjustable phase shifts \cite{7332961}. Compared with these previous works, TFPSPs can exploit the channel sparsity along triple beams. By scheduling ${\phi_u}$ and ${\varphi_u}$ simultaneously, TFPSPs can make the non-overlapping among different equivalent channels much easier. In this case, more UTs can be accomodated by the system without degrading the MSE performance of the channel estimation. Meanwhile, with only a few elements of ${ {\mathbfcal W}}_u$ dominating, even the same TFPSP can be scheduled to UTs with no overlap in the TB domain equivalent channels. Moreover, TFPSPs help to eliminate inter-UT interference within a limited pilot overhead, which positively influences the subsequent transmission design. As shown in \cite{you2015pilot}, achievable spectral efficiency is closely related to pilot overhead and channel estimation accuracy. Therefore, employing TFPSPs can enhance spectral efficiency of the system.

	\subsection{Pilot Scheduling}
	

	Based on the above analysis, scheduling TFPSPs to satisfy the condition (\ref{condition}) can achieve the optimal channel estimation performance. In practical systems, as the number of UTs increases, (\ref{condition}) may not always be well satisfied. However, we can still follow such an idea  and schedule TFPSPs for different UTs to suppress the interference between any two different UTs as much as possible.

	
	To measure the overlap degree of the equivalent TB domain channels of any two different UTs, the following parameter is introduced
	\begin{equation}\label{eta}
		\eta \left( {{{\mathbfcal A}},{{\mathbfcal B}}} \right) \buildrel \Delta \over = \frac{{ {\sum\nolimits_{i,j,k} {{{\left[ {{\mathbfcal A} \odot {\mathbfcal B}} \right]}_{i,j,k}}} } }}{{\sqrt {\sum\nolimits_{i,j,k} {\left[ {\mathbfcal A} \right]_{i,j,k}^2} } \sqrt {\sum\nolimits_{i,j,k} {\left[ {\mathbfcal B} \right]_{i,j,k}^2} } }},
	\end{equation}
	where all the elements in ${\mathbfcal A}$ and ${\mathbfcal B}$ are non-negative real numbers as both ${\mathbfcal A}$ and ${\mathbfcal B}$ stand for power distributions. From Cauchy-Schwarz inequality, $0 \le {\eta \left( {{{\mathbfcal A}},{{\mathbfcal B}}} \right)} \le 1$. ${\eta \left( {{{\mathbfcal A}},{{\mathbfcal B}}} \right)} = 0$ when ${{\mathbfcal A}} \odot {{\mathbfcal B}} = {\bf{0}}$, and ${\eta \left( {{{\mathbfcal A}},{{\mathbfcal B}}} \right)} = 1$ when ${{\mathbfcal A}}$ is a scaled version of ${{\mathbfcal B}}$. Since for $\forall u,u' \in {{\cal U} }$, $u \ne u'$, the elements in ${\mathbfcal W}_u$ and ${\mathbfcal W}_{u'}$ are all non-negative, $\eta \left( {{{\mathbfcal W}_{u,{L_{{\phi _u}}},{L_{{\varphi _u}}}}},{{\mathbfcal W}_{u',{L_{{\phi _{u'}}}},{L_{{\varphi _{u'}}}}}}} \right) = 0$ if and only if  (\ref{condition}) holds.  
	
	\begin{breakablealgorithm}
		
		\caption{Pilot Scheduling Algorithm with TFPSPs} %
		
		\label{algorithm1}
		\begin{algorithmic}[1]
			\footnotesize
			\Require
			The UT set ${{\cal U} }$; ${{\mathbfcal W}_u}$ for $\forall u \in {\cal U} $; $G\left( {{\cal U},{\cal D}} \right)$;  $\gamma$;
			\Ensure
			The phase shift factors ${{\varphi _u}}$ and ${{\phi _u}}$ for $\forall u \in {\cal U} $;\\
			\textbf{Step 1: DSatur UT grouping} 
			\State Initialization: $\forall u \in {\cal U}$,  ${c_u} = 0$; 
			\State Start with the vertex $u_0$ with the most connected vertices and color it, i.e., ${c_{u_0}} = 1$;
			\While {$\exists {c_u} = 0$} 
			\State Select the uncolored vertex $u$ with the most colors in its neighborhood as the current vertex;
			\State Use ${\cal T}_u$ as the set of color labels in the $u$-th vertex's neighborhood, and color $u$ as ${c_u} = \mathop {\min }\limits_{x \in {{\mathbb{N}}_ + },x \notin {{\cal T}_u}} x$
			\EndWhile
			\State $C = \mathop {\max }\limits_{u \in {\cal U}} {c_u}$ and divide $\cal U$ into ${\cal U}_0$, ${\cal U}_1$,$\cdots$, ${\cal U}_{C-1}$ according to the colored $G\left( {{\cal U},{\cal D}} \right)$.
			\For {$i=0,1,\cdots,C-1$} 
			\State ${{\mathbfcal W}_{{{\cal U}_i}}} = \sum\limits_{u \in {{\cal U}_i}} {{{\mathbfcal W}_u}} $;
			\EndFor\\
			\textbf{Step 2: TFPSP scheduling}
			\State Initialization: the set of scheduled UT groups ${{\cal U} ^{{\rm{sch}}}} = \{0\}$, the set of unscheduled UT ${{\cal U} ^{{\rm{uns}}}} = \{0,1,\cdots,C-1\} \backslash {{\cal U} ^{{\rm{sch}}}}$, and ${\phi _{{{\cal U}_0}}} = {\varphi _{{{\cal U}_0}}} = 0$;
			
			\For {each $i \in {{{\cal U} ^{{\rm{uns}}}}}$} 
			\State Search for ${{\varphi}}$, ${{\phi}}$ satisfying \scalebox{0.8}{$\eta \left( {{{{ {\mathbfcal W}}}_{{\cal U}_i,{L_{\varphi}},{L_{\phi}}}},\!\!\sum\limits_{i' \in {{\cal U} ^{{\rm{sch}}}}}^{}\!\!\! {{{{{\mathbfcal W}}}_{{\cal U}_{i'},L_{\varphi _{{\cal U}_{i'}}},L_{\phi _{{\cal U}_{i'}}}}}} } \right) \le \gamma$} and ${\varphi _{{\cal U}_{i}}} = \varphi $, ${\phi _{{\cal U}_{i'}}} = \phi $;
			\State If the phase shift factors cannot be found in the former step, adopt 
			\begin{align}
				\resizebox{0.8\hsize}{!}
				{$\left\{ {{{\varphi _{{\cal U}_{i}}}},{{\phi _{{\cal U}_{i}}}}} \right\} = \mathop {\arg \min }\limits_{\left\{ {\varphi ,\phi } \right\}} \eta \left( {{{{{\mathbfcal W}}}_{{\cal U}_i,{L_\varphi},{L_\phi} }},\sum\limits_{i' \in {{\cal U} ^{{\rm{sch}}}}}^{} {{{{{\mathbfcal W}}}_{{\cal U}_{i'},L_{\varphi _{{\cal U}_{i'}}},L_{\phi _{{\cal U}_{i'}}}}}} } \right)$};\notag
			\end{align}
			\State For each $u \in {{\cal U}_i}$, ${\varphi _u} = {\varphi _{{{\cal U}_i}}}$, ${\phi _u} = {\phi _{{{\cal U}_i}}}$;
			\State Update ${{\cal U} ^{{\rm{sch}}}} \leftarrow {{\cal U} ^{{\rm{sch}}}} \cup \{ i\} $, ${{\cal U} ^{{\rm{uns}}}} \leftarrow {\cal U} \backslash {{\cal U} ^{{\rm{sch}}}}$.
			\EndFor

		\end{algorithmic}
	\end{breakablealgorithm}


With the parameter defined in (\ref{eta}), the TFPSP scheduling problem can be formulated as
\begin{equation}\label{pro}
	\!\!\!\!\!\!\!\!	\scalebox{0.96}{$\mathop {\min }\limits_{\left\{ {{\phi _u},{\varphi _u}:u \in {\cal U}} \right\}} {\rm{   }}\!\!\sum\limits_{u \in {\cal U}} {\sum\limits_{u' \in {\cal U}\backslash \{ u\} }^{}\!\!\!\! {\eta\!\left(  {{{\mathbfcal W}_{u,{L_{{\phi _u}}},{L_{{\varphi _u}}}}},\!{{\mathbfcal W}_{u',{L_{{\phi _{u'}}}},{L_{{\varphi _{u'}}}}}}} \right)} }.$}
\end{equation}
This is a combinational optimization problem, for which we propose a simple approach to find a near-optimal solution. We begin by introducing a threshold value $\gamma$. If the overlap degree between the TB domain channels of two UTs is less than $\gamma$, we can approximate the two UTs as non-overlapping in the TB domain, allowing them to reuse the same TFPSP. Based on this criterion, we classify all UTs into distinct groups denoted as ${{\cal U}_{0}},{{\cal U}_{1}}, \cdots $. In each group, the UTs have non-overlapping TB domain channels and can therefore share the same TFPSP. Then we schedule each UT group ${\cal U}_{i}$ with a specific TFPSP pair $\left\{ {{\phi _{{{\cal U}_i}}},{\varphi _{{{\cal U}_i}}}} \right\}{\rm{ }}$, enabling seperation among the various UT groups in the TB domain.

Motivated by \cite{7478095}, we adopt the degree of saturation algorithm (DSatur) to divide the UTs into different groups according to their overlap degrees. All the UTs in the system and their mutual overlap degrees can be decribed in an undirected weighted graph  $G\left( {{\cal U},{\cal D}} \right)$, where all the UTs are expressed as the vertices and the edges ${\cal D} = \left\{ {\eta \left( {{{\mathbfcal W}_u},{{\mathbfcal W}_{u'}}} \right)|u,u' \in U,\eta \left( {{{\mathbfcal W}_u},{{\mathbfcal W}_{u'}}} \right)  >  \gamma } \right\}$ of the graph. With the graph, we color the vertices one after another and keep the connected vertices in different colors. 
After that, we use a greedy method to select the phase shift factors for each group. The procedures of the whole pilot scheduling algorithm are shown in Algorithm \ref{algorithm1}, where a smaller $\gamma $ corresponds to a better performance as well as a larger computation overhead.

\section{Channel Estimation with IGA}\label{section4}

In this section, we introduce a tensor-based information geometry approach (IGA)  for the estimation of the TB domain channel tensor. Specifically, we reformulate the channel estimation problem as a statistical inference problem and extend the IGA proposed in \cite{9910031} and \cite{RIGA} to the tensor version for estimating the TB domain channel. Additionally, we utilize the specific structure of the beam matrices to simplify the tensor-based IGA process and complete channel prediction.

\subsection{TB Domain Channel Tensor Estimation with IGA}

With TFPSPs properly scheduled, the MMSE estimate in (\ref{estimateHtb}) represents the optimal estimate of  $ {\mathbfcal H}_u^{{\rm{TB}}}$. However, the complexity of calculating $\hat {\mathbfcal H}_u^{{\rm{TB}}}$ in (\ref{estimateHtb}) for all $u\in \cal{U}$ is ${\cal O}\left( {{{\left( {M{K}{N_{\rm{p}}}} \right)}^3} + {{\left( {M{K}{N_{\rm{p}}}} \right)}^2}{N_\upvartheta }{N_\uptau }{N_\upnu }U} \right)$ \cite{tensor2017Si,opentensor}, rendering it impractical in massive MIMO-OFDM systems. 
To tackle this challenge, we convert the TB domain channel estimation problem into a statistical inference problem and extend the IGA introduced in \cite{9910031} and \cite{RIGA} to our case with the TB-based channel model. This approach offers a much simpler means of obtaining ${\mathbfcal H}_u^{{\rm{TB}}}$.

Before introducing IGA, we define ${{\mathbfcal H}^{{\rm{TB}}}} \buildrel \Delta \over = \displaystyle\sum\limits_{u = 0}^{U - 1}\!\! {{\mathbfcal H}_{u,{L_{{\phi _u}}},{L_{{\varphi _u}}}}^{{\rm{TB}}}}$
as the sum of the equivalent TB domain channel tensors of all the UTs. With ${\mathbfcal H}_{}^{{\rm{TB}}}$, the received signal in (\ref{rec1}) can also be expressed as
\begin{align}\label{recA1}
	{{\mathbfcal Y}_T^{\rm{p}}} &= \sqrt {{\sigma ^{\rm{p}}}} {{\mathbfcal X}_T}{ * _3}{\mathbfcal V}_T^{\rm{p}}{ * _3}{\mathbfcal H}_{}^{{\rm{TB}}} + {\mathbfcal Z}= {{{\mathbfcal A}}}{ * _3} {\mathbfcal H}_{}^{{\rm{TB}}} + {\mathbfcal Z},
\end{align}
where ${{\mathbfcal X}_T}\in {{\mathbb{C}}^{M \times {K} \times {N_{\rm{p}}} \times M \times {K} \times {N_{\rm{p}}}}}$ is a pseudo-diagonal tensor, determined by the basic pilot sequences ${{\bf{x}}^{\rm{f}}}$ and ${{\bf{x}}^{\rm{t}}}$ as ${\left[ {{{\mathbfcal X}_T}} \right]_{m,{k},{n},m,{k},{n}}} ={\left[ {{{\bf{x}}^{\rm{f}}}} \right]_k}{\left[ {{{\bf{\Gamma }}_{{N_{\rm{p}}},{n_T}}}{\bf{x}}_{}^{\rm{t}}} \right]_n}$, and ${{ {\mathbfcal A}}} \buildrel \Delta \over = \sqrt {{\sigma ^{\rm{p}}}} {{\mathbfcal X}_T}{ * _3}{\mathbfcal V}_T^{\rm{p}}\in {{\mathbb{C}}^{M \times {K} \times {N_{\rm{p}}} \times {N_{\upvartheta}} \times {N_{\uptau}} \times {N_{\upnu}}}}$. Accordingly, the TB domain power distribution of ${\mathbfcal H}_{}^{{\rm{TB}}}$ equals the sum of the equivalent TB domain power distributions of all the UTs, i.e., ${\mathbfcal W} \buildrel \Delta \over = {\mathbb{E}}\left\{ {{{\mathbfcal H}^{\rm{TB}}} \odot {{\left( {{{\mathbfcal H}^{\rm{TB}}}} \right)}^*}} \right\} = \sum\limits_{u = 0}^{U - 1} {{{\mathbfcal W}_{u,{L_{{\phi _u}}},{L_{{\varphi _u}}}}}}. $ With (\ref{recA1}), the MMSE estimate of ${{{\mathbfcal H}^{\rm{TB}}}} $ can be expressed as \cite{pandey2021multi}
\begin{equation}\label{goon2}
	\scalebox{0.98}{${{\hat {\mathbfcal H}}^{{\rm{TB}}}} = {\mathbfcal R}_{}^{{\rm{TB}}}{ * _3}{\mathbfcal A}_3^{\rm{H}}{ * _3}{\left( {{\mathbfcal A}{ * _3} {\mathbfcal R}_{}^{{\rm{TB}}}{ * _3}{\mathbfcal A}_3^{\rm{H}} + {{\mathbfcal I}_{M,K,{N_{\rm{p}}}}}} \right)^{ - 1}}{ * _3}{{\mathbfcal Y}_T^{\rm{p}}}$},
\end{equation}
where $ {\mathbfcal R}^{{\rm{TB}}}$$\in{{\mathbb{C}}^{{N_\upvartheta } \times {N_\uptau } \times {N_\upnu } \times {N_\upvartheta } \times {N_\uptau } \times {N_\upnu }}}$ is  pseudo-diagonal, and ${\left[ {{\mathbfcal R}^{\rm{TB}}} \right]_{{n_1},{n_2},{n_3},{n_1},{n_2},{n_3}}}\!\! = \!\!{\left[ {{\mathbfcal W}} \right]_{{n_1},{n_2},{n_3}}}$. Meanwhile, we can also rewritten $\hat{\mathbfcal H}_u^{{\rm{TB}}}$ in (\ref{estimateHtb}) with $\mathbfcal A$ as
\begin{align}
	\scalebox{0.98}{$\hat {\mathbfcal H}_u^{{\rm{TB}}} =$} &\scalebox{0.98}{$\sqrt {{\sigma _{\rm{p}}}} {\mathbfcal R}_u^{{\rm{TB}}}{ * _3}\left( {{{\mathbfcal X}_{u,T}}{ * _3}{\mathbfcal V}_T^{\rm{p}}} \right)_3^{\rm{H}}$}\notag\\
	&\scalebox{0.98}{${ * _3}{\left( {{\mathbfcal A}{ * _3} {\mathbfcal R}_{}^{{\rm{TB}}}{ * _3}{\mathbfcal A}_3^{\rm{H}} + {{\mathbfcal I}_{M,K,{N_{\rm{p}}}}}} \right)^{ - 1}}{ * _3}{{\mathbfcal Y}_T^{\rm{p}}}.$}\label{goon1}
\end{align}
From (\ref{goon2}) and (\ref{goon1}), once $\hat {\mathbfcal H}^{{\rm{TB}}}$ is acquired, the MMSE estimate of  ${\mathbfcal H}_u^{{\rm{TB}}}$ for $\forall u \in { \cal U} $ can be calculated by
\begin{equation}\label{goon3}
	\!\!\!\!\!\!\hat{\mathbfcal H}_u^{{\rm{TB}}} = \sqrt {{\sigma _{\rm{p}}}} {\mathbfcal R}_u^{{\rm{TB}}}{ * _3}\left( {{{\mathbfcal X}_{u,T}}{ * _3}{\mathbfcal V}_T^{\rm{p}}} \right)_3^{\rm{H}}{ * _3}{\left( { {\mathbfcal R}_{}^{{\rm{TB}}}} \right)^{\dag}}{ * _3}\hat{\mathbfcal H}_{}^{{\rm{TB}}}.
\end{equation}
Therefore, we can transform the original estimation of ${\mathbfcal H}_u^{{\rm{TB}}}$ for $\forall u \in { \cal U} $ to the estimation of ${\mathbfcal H}^{{\rm{TB}}}$.

We can regard  ${\mathbfcal H}_{}^{{\rm{TB}}}$ as the TB domain channel tensor of a single UT with the TB domain power distribution $ {\mathbfcal W}$. The MMSE estimation of ${{\mathbfcal H}^{{\rm{TB}}}}$ is equivalent to calculating the expectation of the posterior PDF $p\left( {{{\mathbfcal H}^{{\rm{TB}}}}|{{\mathbfcal Y}_T^{\rm{p}}}} \right)$, i.e., ${\hat{\mathbfcal H}}_{}^{{\rm{TB}}} = {{\mathbb{E}}_{p\left( {{\mathbfcal{H}}_{}^{{\rm{TB}}}{\rm{|}}{{\mathbfcal{Y}}_T^{\rm{p}}}} \right)}}\left\{ {{\mathbfcal{H}}_{}^{{\rm{TB}}}} \right\}$. To simplify the representations, we use vector ${\bf{b}}$ instead of $\left( {{n_\upvartheta },{n_\uptau },{n_\upnu }} \right)$ as the TB domain index. Then, we have the following posterior PDF:
\begin{align}\label{tensorpdf}
	\scalebox{0.95}{$\!\!\!\!p\left( {{{\mathbfcal H}^{{\rm{TB}}}}|{{\mathbfcal Y}_T^{\rm{p}}}} \right)$}
	=\scalebox{0.95}{$\! \exp \!\left\{ {\! -\! \displaystyle\sum\limits_{{\bf b} \in {{\cal S}_{{\rm{nz}}}}} \!\!\!{\frac{{\left[ {{\mathbfcal H}_{}^{{\rm{TB}}}} \right]_{\bf {b}}^*\left[ {{\mathbfcal H}_{}^{{\rm{TB}}}} \right]_{\bf b}^{}}}{{{{\left[ {\mathbfcal W} \right]}_{\bf b}}}}}  + d + {\psi _p}} \right\},$}
\end{align}
where ${{{\cal S}_{{\rm{nz}}}}}$ represents the index set of non-zero elements in ${\mathbfcal W}$, and ${{\psi_p}}$ denotes the normalized factor, which is independent of any element in ${\mathbfcal H}_{}^{{\rm{TB}}}$. Since ${\mathbfcal H}_{}^{{\rm{TB}}}$ is determined by ${\mathbfcal H}_{u}^{{\rm{TB}}}$ and $({{{\phi _u}}},{{{\varphi _u}}})$ for all the $u \in \cal U$, ${{{\cal S}_{{\rm{nz}}}}}$ can then be acquired from ${\cal S}_u$ and $({L_{{\phi _u}}},{L_{{\varphi _u}}})$ of all the UTs.
With $A \buildrel \Delta \over = M{K}{N_{\rm{p}}}$, ${a_{\rm{s}}} = {\left\langle a-1 \right\rangle _{M}} $, ${a_{\rm{f}}} = {\left\langle {\left\lfloor {(a-1)/M} \right\rfloor } \right\rangle _{{K}}}$ and ${a_{\rm{t}}} = \left\lfloor {(a-1)/\left( {M{K}} \right)} \right\rfloor $, the term $d$ in (\ref{tensorpdf}) is defined as $d = \sum\limits_{a = 1}^{A} {d_{a}} $, where
\begin{align}
	\scalebox{1}{${d_a}$} =  &\scalebox{0.98}{$- \frac{1}{{{\sigma _z}}}\left( {{{\left| {{{\left[ {\mathbfcal{A}} \right]}_{{a_{\rm{s}}},{a_{\rm{f}}},{a_{\rm{t}}},:,:,:}}{*_3}{\mathbfcal{H}}_{}^{{\rm{TB}}}} \right|}^2} - } \right.$}\notag\\
	&\scalebox{0.98}{$\left. {2{\rm{Re}}\left\{ {\left[ {{{\mathbfcal{Y}}_T^{\rm{p}}}} \right]_{{a_{\rm{s}}},{a_{\rm{f}}},{a_{\rm{t}}}}^*{{\left[ {\mathbfcal{A}} \right]}_{{a_{\rm{s}}},{a_{\rm{f}}},{a_{\rm{t}}},:,:,:}}{*_3}{\mathbfcal{H}}_{}^{{\rm{TB}}}} \right\}} \right).$}
\end{align}
The term $d$ captures the interactions among different elements in ${\mathbfcal H}{}^{{\rm{TB}}}$. It is commonly referred to as the interaction item (or crossing terms) \cite{amari2000methods}. 
We define the tensor ${{\mathbfcal T}_h} \in {{\mathbb{C}}^{{N_\upvartheta } \times {N_\uptau } \times {N_\upnu } \times 3}}$, where ${\left[ {{{\mathbfcal T}_h}} \right]_{:,:,:,1}} = {{\mathbfcal H}^{{\rm{TB}}}}$, ${\left[ {{{\mathbfcal T}_h}} \right]_{:,:,:,2}} = {\left( {{{\mathbfcal H}^{{\rm{TB}}}}} \right)^*}$ and ${\left[ {{{\mathbfcal T}_h}} \right]_{:,:,:,3}} = {{\mathbfcal H}^{{\rm{TB}}}} \odot {\left( {{{\mathbfcal H}^{{\rm{TB}}}}} \right)^*}$, serving as the sufficient statistic of ${{\mathbfcal H}^{{\rm{TB}}}}$. Additionally, we define ${\mathbfcal G} \in {{\mathbb{C}}^{{N_\upvartheta } \times {N_\uptau } \times {N_\upnu } \times 3}}$, where ${\left[ {\mathbfcal G} \right]_{:,:,:,1}} = {\left[ {\mathbfcal G} \right]_{:,:,:,2}}= \bf 0 $ and ${\left[ {\mathbfcal G} \right]_{:,:,:,3}} = {\mathbfcal W}^\dag $. With these definitions, the posterior PDF in (\ref{tensorpdf}) can be further expressed as
\begin{equation}\label{postpdf}
	p\left( {{{\mathbfcal H}^{{\rm{TB}}}}|{{\mathbfcal Y}_T^{\rm{p}}}} \right) \!\! = \!\exp \left\{ { - {{\mathbfcal T}_h}{*_4}{\mathbfcal G} + d + {\psi _p}} \right\}.
\end{equation}
and our objective is to find an alternative PDF with the same expectation as $p\left( {{{\mathbfcal H}^{{\rm{TB}}}}|{{\mathbfcal Y}_T^{\rm{p}}}} \right)$.


Let us define a manifold of PDFs ${\cal M}_0$, called the objective manifold (OBM). Any PDF in ${\cal M}_0$ can be expressed in the form of
\begin{align}\label{p0}
	{p_0}\left( {{{\mathbfcal H}^{{\rm{TB}}}},{{{\mathbfcal L}_0}}} \right) &=  \exp \left\{ { - {{\mathbfcal T}_h}{*_4}{\mathbfcal G} + {{\mathbfcal T}_h}{*_4}{\mathbfcal L}_0 + {\psi _{0,{\mathbfcal L}_0}}} \right\},
\end{align}
where ${\psi _{0,{\mathbfcal L}_0}}$ is the norminalized factor and ${\mathbfcal L}_0$ is called the natural parameter (NP).  ${\mathbfcal L}_0$ is an ${{N_\upvartheta } \times {N_\uptau } \times {N_\upnu } \times 3}$ tensor including statistical means and variances, which are called the first-order NP (FONP) and the second-order NP (SONP), respectively. Specifically, ${[{{\mathbfcal L}_0}]_{:,:,:,1}}$$= [{{\mathbfcal L}_0}]_{:,:,:,2}^* \in {{\mathbb{C}}^{{N_\upvartheta } \times {N_\uptau } \times {N_\upnu }}}$  represents the FONP, and ${[{{\mathbfcal L}_0}]_{:,:,:,3}}  \in {{\mathbb{R}}^{{N_\upvartheta } \times {N_\uptau } \times {N_\upnu }}}$ represents the SONP. Meanwhile, the positions of non-zero elements in the FONP and SONP are the same as $\mathbfcal W$. For convience, we use ${{\mathbfcal D}_0}  \buildrel \Delta \over =  {[{{\mathbfcal L}_0}]_{:,:,:,1}}$$= [{{\mathbfcal L}_0}]_{:,:,:,2}^*$ as the FONP and ${{\mathbfcal F}_0}  \buildrel \Delta \over =  {[{{\mathbfcal L}_0}]_{:,:,:,3}}$ as the SONP in the following.

Compared with the PDF in (\ref{postpdf}), ${p_0}$ in (\ref{p0}) is a Gaussian distribution without iteraction terms, and it can be factorized as the product of its marginals, i.e., ${p_0}\left({{{\mathbfcal H}^{{\rm{TB}}}},{{{\mathbfcal L}_0}}} \right) = \prod\limits_{{\bf{b}} \in {{\cal S}_{{\rm{nz}}}}} {{p_0}\left( {{\left[ {{\mathbfcal H}_{}^{{\rm{TB}}}} \right]_{\bf{b}}},{{{\mathbfcal L}_0}}} \right)} $. In this case, the expectation of ${{\left[ {{{\mathbfcal H}^{{\rm{TB}}}}} \right]}_{{\bf b}}}$ for each ${\bf{b}} \in {{\cal S}_{{\rm{nz}}}}$ can be acquired easily from ${p_0}\left( {{{\mathbfcal H}^{{\rm{TB}}}},{{{{\mathbfcal L}_0}}}} \right) $. For all the ${p_0}\left( {{{\mathbfcal H}^{{\rm{TB}}}},{{{{\mathbfcal L}_0}}}} \right) \in {{\cal M}_0}$, we define the distribution that minimizes the Kullback-Leibler (KL) divergence from $p\left( {{{\mathbfcal H}^{{\rm{TB}}}}|{{\mathbfcal Y}_T^{\rm{p}}}} \right)$ to ${\cal M}_0$ as follows 
\begin{equation}\label{p0proj}
{p_0}\left( {{{\mathbfcal H}^{{\rm{TB}}}},{{{\mathbfcal L}_0^{{\rm{proj}}}}}} \right) \!\!=\!\! \mathop {\arg \min }\limits_{{p_0}\left( {{{\mathbfcal H}^{{\rm{TB}}}},{{{\mathbfcal L}_0}}^{}} \right)} {{\mathbb{E}}_{p\left( {{{\mathbfcal H}^{{\rm{TB}}}}|{{\mathbfcal Y}_T^{\rm{p}}}} \right)}}\!\!\left\{ {\frac{{p\left( {{{\mathbfcal H}^{{\rm{TB}}}}|{{\mathbfcal Y}_T^{\rm{p}}}} \right)}}{{{p_0}\left( {{{\mathbfcal H}^{{\rm{TB}}}},{{{\mathbfcal L}_0}}^{}} \right)}}} \right\},
\end{equation}
and ${p_0}\left( {{{\mathbfcal H}^{{\rm{TB}}}},{{{\mathbfcal L}_0^{{\rm{proj}}}}}} \right)$ is also referred as the \emph{m-projection} of ${p\left( {{{\mathbfcal H}^{{\rm{TB}}}}|{{\mathbfcal Y}_T^{\rm{p}}}} \right)}$ onto the OBM \cite{9910031}. The \emph{m-projection} is unique and preserves the first and second moments of the original distribution \cite{ikeda2004information}. Therefore, we have ${{\mathbb{E}}_{p\left( {{\mathbfcal{H}}_{}^{{\rm{TB}}}{\rm{|}}{{\mathbfcal{Y}}_T^{\rm{p}}}} \right)}}\left\{ {{\mathbfcal{H}}_{}^{{\rm{TB}}}} \right\}$ = ${{\mathbb{E}}_{{p_0}\left( {{{\mathbfcal H}^{{\rm{TB}}}},{{{\mathbfcal L}_0^{{\rm{proj}}}}}} \right)}}\left\{ {{{\mathbfcal H}^{\rm{TB}}}} \right\} $, and ${\hat{\mathbfcal H}}_{}^{{\rm{TB}}}$ can be computed from ${p_0}\left( {{{\mathbfcal H}^{{\rm{TB}}}},{{{\mathbfcal L}_0^{{\rm{proj}}}}}} \right)$ easily. The objective of IGA is to find such an \emph{m-projection}. The tensor-based IGA procedure for TB domain channel tensors is provided below, while additional details about IGA are available in \cite{9910031, RIGA, yang2024simplified2, amari2016information}.

Since IGA finds the corresponding \emph{m-projection} to approximate $p\left( {{{\mathbfcal H}^{{\rm{TB}}}}|{{\mathbfcal Y}_T^{\rm{p}}}} \right)$, the interaction item $d$ in $p\left( {{{\mathbfcal H}^{{\rm{TB}}}}|{{\mathbfcal Y}_T^{\rm{p}}}} \right)$ is approximated by ${{\mathbfcal T}_h}{*_4}{\mathbfcal L}_0^{{\rm{proj}}}$. However, directly calculating ${\mathbfcal L}_0^{{\rm{proj}}}$ by solving (\ref{p0proj}) may be complicated  \cite{9910031}. Instead, we construct $A$ auxiliary manifolds (AMs), denoted as ${{\cal M}_a}$, where $a \in \{ 1,2, \cdots ,A\} $, to solve our problem. All the PDFs in ${{\cal M}_a}$ can be expressed in the form of 
\begin{equation}\label{pn}
{p_a}\left( {{{\mathbfcal H}^{{\rm{TB}}}},{{{{\mathbfcal L}_a}}}} \right) = \exp \left\{ { - {{\mathbfcal T}_h}{*_4}{\mathbfcal G} + {{\mathbfcal T}_h}{*_4}{{\mathbfcal L}_a} + {d_{a}} + {\psi _{a,{{\mathbfcal L}_a}}}} \right\},
\end{equation}
where the interaction item is ${d_a}$, and ${\psi _{a,{{\mathbfcal L}_a}}}$ is the normalized factor. Besides, the NP ${{\mathbfcal L}_a}\in {{\mathbb{C}}^{{N_\upvartheta } \times {N_\uptau } \times {N_\upnu } \times 3}}$ should satisfy the same condition as ${{\mathbfcal L}_0}$ in (\ref{p0}) and we let $ {\mathbfcal D}_a\buildrel \Delta \over ={\left[ {\mathbfcal L}_a \right]_{:,:,:,1}} $ and ${\mathbfcal F}_a\buildrel \Delta \over ={\left[ {\mathbfcal L}_a \right]_{:,:,:,3}}$. Compared with (\ref{tensorpdf}), in the format of (\ref{pn}), $\sum\limits_{a' \ne a} {{d_{a'}}}$ in (\ref{tensorpdf}) is replaced by ${{\mathbfcal T}_h}{*_4}{{\mathbfcal L}_a}$. Each ${p_a}\left( {{{\mathbfcal H}^{{\rm{TB}}}},{{{{\mathbfcal L}_a}}}} \right)$ can also be projected onto ${\cal M}_{0}$ to obtain its corresponding \emph{m-projection}, written as ${p_0}\left( {{{\mathbfcal H}^{{\rm{TB}}}},{{{{\mathbfcal L}_{0a}}}}} \right)$, where $ {\mathbfcal D}_{0a}\buildrel \Delta \over ={\left[ {\mathbfcal L}_{0a} \right]_{:,:,:,1}} $ and ${\mathbfcal F}_{0a}\buildrel \Delta \over ={\left[ {\mathbfcal L}_{0a} \right]_{:,:,:,3}}$. 
This \emph{m-projection} also  keeps the expectation invariant, i.e., ${{\mathbb{E}}_{{p_a}\left( {{{\mathbfcal H}^{{\rm{TB}}}},{{{{\mathbfcal L}_a}}}} \right)}}\left\{ {{{\mathbfcal H}^{{\rm{TB}}}}} \right\}$${\rm{ = }}$ ${{\mathbb{E}}_{{p_a}\left( {{{\mathbfcal H}^{{\rm{TB}}}},{{{{\mathbfcal L}_{0a}}}}} \right)}}\left\{ {{{\mathbfcal H}^{{\rm{TB}}}}} \right\}$, and it can be regarded as approximating ${p_a}\left( {{{\mathbfcal H}^{{\rm{TB}}}},{{{{\mathbfcal L}_a}}}} \right)$ in (\ref{pn}) with ${p_0}\left( {{{\mathbfcal H}^{{\rm{TB}}}},{{{{\mathbfcal L}_{0a}}}}} \right)$. Accordingly, $d_{a}$ in (\ref{pn}) can be approximated by
\begin{equation}\label{dna}
{d_a} \approx {\mathbfcal T}_h{ * _4}\left( {{{\mathbfcal L}_{0a}} - {{\mathbfcal L}_a}} \right) = {\mathbfcal T}_h{ * _4}{{\widetilde {\mathbfcal L}}_a},
\end{equation}
where ${{\widetilde {\mathbfcal L}}_a}  \buildrel \Delta \over =  {{\mathbfcal L}_{0a}} - {{\mathbfcal L}_a}$. Then the interaction item $d$ in (\ref{postpdf}) can be approximated by $d = \sum\limits_{a = 1}^{A} {{d_a}}  \approx \sum\limits_{a = 1}^{A} {{\mathbfcal T}_h{ * _4}{{\widetilde {\mathbfcal L}}_a}} $.

Based on the above analysis, IGA solves the \emph{m-projection} problem in an iterative way. By comparing (\ref{p0}) and (\ref{pn}), we can use the approximation in (\ref{dna}) and update ${{{\mathbfcal L}_a}}$ by $\sum\limits_{a \ne a'} {{{\widetilde {\mathbfcal L}}_{a'}}} $, ${{{\mathbfcal L}_0}}$ by $\sum\limits_{a = 1}^{A} {{{\widetilde {\mathbfcal L}}_a}} $ during each iteration. Thus, in the whole process of IGA, we begin by initializing the NPs ${\mathbfcal L}_{\rm{0}}^{\rm{0}}{\rm{,}}  {\mathbfcal L}_1^{\rm{0}},  \cdots , {\mathbfcal L}_{A}^{\rm{0}}$, and in the $t$-th iteration, the following steps are performed to update the parameters:
\begin{subequations}\label{IGAstep}
\begin{align}
	\!\!\!\!	{\mathbfcal L}_a^{t + 1} &= \alpha \sum\limits_{a' \ne a} {\widetilde {\mathbfcal L}{{_{a'}^t}_{}}}  + \left( {1 - \alpha } \right){\mathbfcal L}_a^t, \quad\!\!\! a = 1,2, \cdots ,A,\\
	\!\!\!\!	{\mathbfcal L}_0^{t + 1} &= \alpha \sum\limits_{a = 1}^{A} {\widetilde {\mathbfcal L}{{_{a}^t}_{}}}  + \left( {1 - \alpha } \right){\mathbfcal L}_0^t,
\end{align}
\end{subequations}
where $0 < \alpha  \le 1$ is the damping factor \cite{ikeda2004information}. It has been shown in \cite{yang2024simplified2} that IGA can converge to a fixed point by properly choosing the damping factor. At the fixed point of IGA, the expectation of ${p_0}$ is equivalent to the MMSE estimation \cite[Theorem 2]{9910031}. From (\ref{IGAstep}), the NP ${{\mathbfcal L}_{0a}}$ for $a=1,2,\cdots,A$ need to be calculated in each iteration. We can refer to \cite[Theorem 1]{9910031} and generate the parameters in the similar way.

However, the above procedure of IGA can be further simplified. According to \cite[Theorem 1, Theorem 2]{RIGA}, under the conditions of sufficiently large $A$, the NPs ${\mathbfcal L}_{{a}}^{t}$ in the $t$-th iteration for any $a \in \left\{ {1,2, \cdots ,A} \right\}$ can be well approximated by their arithmetic mean ${{\mathbfcal L}^t} $, i.e., ${{\mathbfcal L}^t} = \frac{1}{A}\sum\limits_{a = 1}^{A} {{\mathbfcal L}_a^t} $. In this context, for any $a, a' \in \left\{ {1,2, \cdots ,A} \right\}$ and $a \neq a'$, we have ${\mathbfcal L}_a^t = {\mathbfcal L}_{a'}^t = {{\mathbfcal L}^t}$. Consequently, the iterative step in (\ref{IGAstep}) can be improved to:
\begin{subequations}\label{IGAstep1}
\begin{align}
	{{\mathbfcal L}^{t + 1}} &= \frac{{\alpha \left( {A - 1} \right)}}{A}\sum\limits_{a = 1}^{A} {{\mathbfcal L}_{0a}^t}  + \left( {1 - \alpha A} \right){{\mathbfcal L}^t},\\
	{\mathbfcal L}_0^{t{\rm{ + 1}}}&{\rm{ = }}\alpha \sum\limits_{a = 1}^{A} {{\mathbfcal L}_{0a}^t}  - \alpha {A}{{\mathbfcal L}^{t + 1}} + \left( {1 - \alpha } \right){\mathbfcal L}_0^t.
\end{align}
\end{subequations}
By following the steps in (\ref{IGAstep1}), we can initiate the IGA by setting $A{{\mathbfcal L}^0} = \left( {A - 1} \right){\mathbfcal L}_0^0$, and this results in ${\mathbfcal L}_0^{t} = \frac{A}{{A - 1}}{{{\mathbfcal L}}^{t}}$ in each iteration. Let ${{\mathbfcal D}^{t}} \buildrel \Delta \over = {\left[ {{{\mathbfcal L}^{t}}} \right]_{:,:,:,1}}$ $={\left[ {{{\mathbfcal L}^{t}}} \right]_{:,:,:,2}^*}$ and $ {{\mathbfcal F}^{t}} \buildrel \Delta \over = {\left[ {{{\mathbfcal L}^{t}}} \right]_{:,:,:,3}}$. Leveraging  the conclusion in (\ref{IGAstep1}), we can make a calculation similar to the Appendix C in \cite{RIGA} and generate the parameters in the $t$-th iteration as
\begin{subequations}\label{IGAstep2}
\begin{align}
	\!\!\!\!	\scalebox{0.98}{${{\mathbfcal D}^{t + 1}} $}&= \scalebox{0.98}{$\frac{{\alpha \left( {A - 1} \right)}}{A}{\left( {{\mathbfcal J} - {\gamma ^t}{{\widetilde{\mathbfcal R}}^t}} \right)^\dag }{*_3}\left[ {A{{\mathbfcal D}^t}} + \right.$}\notag\\
	\!\!\!\!	&\scalebox{0.98}{$\left. {{\gamma ^t}{{\mathbfcal A}_3^{\rm{H}}}{*_3}\left( {{{\mathbfcal Y}_T^{\rm{p}}} - {{ {\mathbfcal A}}}{*_3}{{\widetilde{\mathbfcal R}}^t}{*_3}{{\mathbfcal D}^t}} \right)} \right] + \left( {1 - \alpha A} \right){{\mathbfcal D}^t},$}\label{stepa}\\
	\!\!\!\!	\scalebox{0.98}{${{\mathbfcal F}^{t + 1}}$}& \scalebox{0.98}{$= \alpha \left( {A - 1} \right)\left[ {{{\mathbfcal W}^\dag } - {{\left( {{{\mathbfcal P}^t} - {\gamma ^t}{{\mathbfcal P}^t} \odot {{\left( {{{\mathbfcal P}^t}} \right)}^\dag}} \right)}^\dag }} \right]$}\notag\\ 
	\!\!\!\!&\scalebox{0.98}{$ + \left( {1 - \alpha A} \right){{\mathbfcal F}^t},$}
\end{align}
\end{subequations}
where ${{\mathbfcal P}^{t}} \buildrel \Delta \over = {\left( {{{\mathbfcal W}^\dag } - {{\mathbfcal F}^{t}}} \right)^\dag }$, ${\gamma ^t} \buildrel \Delta \over = {\left( {{\sigma _z} + \sum\limits_{{\bf{b}} \in {{\cal S}_{{\rm{nz}}}}} {\left[ {{{\mathbfcal P}^t}} \right]_{\bf{b}}^{}} } \right)^{ - 1}}$, and the tensors ${{{{\widetilde {\mathbfcal R}}}^t}}$,  ${\mathbfcal J} \in {{\mathbb{C}}^{{N_\upvartheta } \times {N_\uptau } \times {N_\upnu } \times {N_\upvartheta } \times {N_\uptau } \times {N_\upnu }}}$ are pseudo-diagonal, satisfying ${\left[ {{{{\widetilde {\mathbfcal R}}}^t}} \right]_{{\bf{b}},{\bf{b}}}} = {\left[ {{{\mathbfcal P}^t}} \right]_{\bf{b}}}$ and ${\left[ {\mathbfcal J} \right]_{{\bf{b,b}}}} = 1$ for all ${\bf{b}} \in {{\cal S}}_{{\rm{nz}}}$, respectively.
Then the NP ${{\mathbfcal L}^{t + 1}}$ can be updated with ${{\mathbfcal D}^{t + 1}}$ and ${{\mathbfcal F}^{t + 1}}$ as ${\left[ {{{\mathbfcal L}^{t + 1}}} \right]_{:,:,:,1}} = {{\mathbfcal D}^{t + 1}}$, ${\left[ {{{\mathbfcal L}^{t + 1}}} \right]_{:,:,:,2}} = {\left( {{{\mathbfcal D}^{t + 1}}} \right)^*}$ and ${\left[ {{{\mathbfcal L}^{t + 1}}} \right]_{:,:,:,3}} = {{\mathbfcal F}^{t + 1}}$. Compared with the iterative steps in (\ref{IGAstep}), (\ref{IGAstep2}) omits the calculation of NPs for different $a$, and the NP ${\mathbfcal L}_0^t$ on the OBM can be directly obtained by ${\mathbfcal L}_0^{t} = \frac{A}{{A - 1}}{{{\mathbfcal L}}^{t}}$. The process of the tensor-based IGA is given in Algorithm \ref{riga}. Meanwhile, we provide a flowchart, as shown in Fig. 3, which details the procedures of estimating ${{{\mathbfcal H}^{{\rm{TB}}}}}$ with the tensor-based IGA and subsequently recovering ${\mathbfcal H}_u^{{\rm{TB}}}$ for each $u \in {\cal U}$.

\begin{figure}[t!]
\centering
\includegraphics[width =180pt]{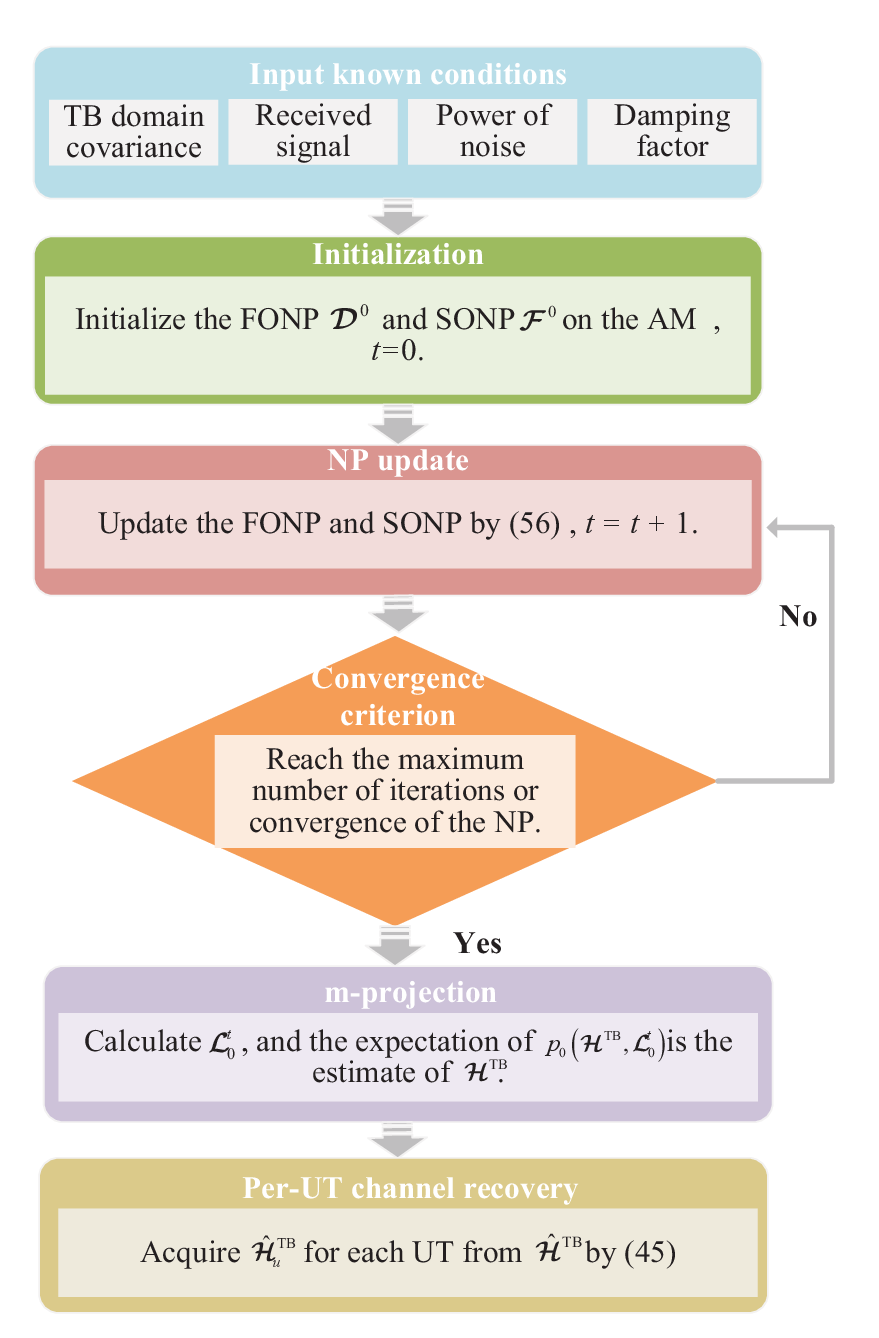}
\caption{Flowchart of the channel estimation framework} 
\label{procedure}
\end{figure}

\begin{breakablealgorithm}

\caption{Tensor-based IGA} %

\label{riga}
\begin{algorithmic}[1]
	\footnotesize
	\Require
	${{\mathbfcal{R}}^{{\rm{TB}}}}$, ${{\mathbfcal{Y}}^{\rm{p}}}$, ${\sigma _z}$ and the maximum iteration number ${t_{\max }}$, the damping factor $\alpha$;
	\Ensure
	The estimate of ${{{\mathbfcal{H}}^{{\rm{TB}}}}}$;
	\State Initialization: $t=0$, ${{\mathbfcal L}^0}$, where $ {{\mathbfcal D}^0} \in {{\mathbb{C}}^{{N_\upvartheta } \times {N_\uptau } \times {N_\upnu }}}$ and  ${{\mathbfcal F}^0} \in {{\mathbb{R}}^{{N_\upvartheta } \times {N_\uptau } \times {N_\upnu }}}$, and the positions of non-zero elements in ${{\mathbfcal D}^0}$ and ${{\mathbfcal F}^0}$ are the same as ${{\mathbfcal W}}$; 
	\Repeat 
	\State Update ${{{\mathbfcal L}}^{t + 1}}$ by (\ref{IGAstep2});
	\State $t=t+1$;
	\Until{$t > {t_{\max }}$ or convergence}
	\State Calculate the corresponding ${\mathbfcal L}_0^{t} $ by ${\mathbfcal L}_0^{t} = \frac{A}{{A - 1}}{{{\mathbfcal L}}^{t}}$;
	\State Estimate  ${{{\mathbfcal{H}}^{{\rm{TB}}}}}$ with ${p_0}\left( {{{\mathbfcal H}^{{\rm{TB}}}},{\mathbfcal L}_0^t} \right)$.

\end{algorithmic}
\end{breakablealgorithm} 
\vspace{-0.4cm}

\subsection{Low-Complexity Implementation of the Tensor-based IGA and Channel Prediction}

The expression in (\ref {IGAstep2}) indicates that, in the process of IGA, the Einstein products between ${{{{\mathbfcal A}}}}$ (or ${\mathbfcal A}_3^{\rm{H}}$) with other tensors occupy most of the computations, and the corresponding computational complexity is ${\cal O}\left( {A\left| {{{\cal S}_{{\rm{nz}}}}} \right|} \right)$. Fortunately, the structure of  ${{{{\mathbfcal A}}}}$ allows us to utilize multi-dimensional FFT to reduce the computational complexity of IGA.


Given any tensor ${\mathbfcal B} \in {\mathbb{C}}^{{N_\upvartheta } \times {N_\uptau } \times {N_\upnu }}$, with ${{\bf{X}}^{\rm{f}}} \buildrel \Delta \over = {\rm{Diag}}\{ {{\bf{x}}^{\rm{f}}}\} $ and ${\bf{X}}_T^{\rm{t}} \buildrel \Delta \over = {\rm{Diag}}\left\{ {{{\bf{\Gamma }}_{{N_{\rm{p}}},{n_T}}}{\bf{x}}_{}^{\rm{t}}} \right\}$, we have
\begin{align}\label{S1b}
\scalebox{0.95}{${{{\mathbfcal A}}}{ * _3}{\mathbfcal B}= \left( {{{\bf{X}}_T^{\rm{t}}}{{\bf{V}}_T^{\rm{t,p}}}} \right){ \times _3}\left( {\left( {{{\bf{X}}^{\rm{f}}}{{\bf{V}}^{\rm{f}}}} \right){ \times _2}\left( {\left( {{{\bf{V}}^{\rm{s}}}} \right){ \times _1}\sqrt {{\sigma ^{\rm{p}}}} {\mathbfcal B}} \right)} \right).$}
\end{align} 
From Section \ref{section2}-C, all the beam matrices can be generated from DFT matrices. Additionally, ${{{\bf{X}}^{\rm{f}}}}$ and ${{\bf{X}}^{\rm{t}}}$ are diagonal matrices. As a result, the complexity of calculating ${{{\mathbfcal A}}}{ * _3}{\mathbfcal B}$ becomes ${\cal O}\left( {{{\bar C}_1}} \right)$, where ${{\bar C}_1}$ $ = $$A\left( {{F_\upvartheta }{F_\uptau }{F_\upnu }{{\log }_2}{N_\upvartheta }} \right. + {F_\uptau }{F_\upnu }{\log _2}{N_\uptau } + \left. {{F_\upnu }{{\log }_2}{N_\upnu }} \right)$. On the other hand, for any ${\mathbfcal C} \in {{\mathbb{C}}^{M \times {K} \times {N_{\rm{p}}}}}$, 
\begin{align}\label{S2}
\!\!\!\!	\scalebox{0.925}{${\mathbfcal A}_3^{\rm{H}}{ * _3}{\mathbfcal C}\!={\left( {{{\bf{X}}_T^{\rm{t}}}{{\bf{V}}_T^{\rm{t,p}}}} \right)^{\rm{H}}}\!{ \times _3}\!\left( {{{\left( {{{\bf{X}}^{\rm{f}}}{{\bf{V}}^{\rm{f}}}} \right)}^{\rm{H}}}\!{ \times _2}\!\left( {{{\left( {{{\bf{V}}^{\rm{s}}}} \right)}^{\rm{H}}}{ \times _1}\sqrt {{\sigma ^{\rm{p}}}} {\mathbfcal C}} \right)} \right).$}
\end{align}
Similar to (\ref{S1b}), ${\mathbfcal A}_3^{\rm{H}}{ * _3}{\mathbfcal C}$ in (\ref{S2}) can also be solved in the complexity of ${\cal O}\left( {{{\bar C}_1}} \right)$.  Then, to implement the iteration of IGA in (\ref{IGAstep2}), the total complexity of IGA is ${\cal O}\left( {{{\bar C}_1}} \right)$. Compared with other  channel estimation algorithms, such as the GAMP algorithm in  \cite{6033942} or  the EPV algorithm in \cite{9351786}, with the complexity of ${\cal O}\left( {A\left| {{{\cal S}_{{\rm{nz}}}}} \right|} \right)$, our proposed algorithm has a lower complexity. With the proposed frame structure, once ${{\mathbfcal H}_u^{{\rm{TB}}}}$ is estimated, we can also predict the SFT domain channel tensor of the data segment in the current $T$-th slot by
\begin{equation}
\!\!\!\!\!\!\!\!\scalebox{0.98}{${\mathbfcal H}_{u,T}^{{\rm{SFT}},{\rm{d}}} = {\left[ {{{\bf{V}}_T^{\rm{t}}}} \right]_{N_{\rm s} - {N_{\rm{b}}} + 1:N_{\rm s},:}}{ \times _3}\left( {{{\bf{V}}^{\rm{f}}}{ \times _2}\left( {{{\bf{V}}^{\rm{s}}}{ \times _1}{\mathbfcal H}_u^{{\rm{TB}}}} \right)} \right).$}
\end{equation}
With the special DFT structures of ${{\bf{V}}^{\rm{s}}}$, ${{\bf{V}}^{\rm{f}}}$, and ${{\bf{V}}_T^{\rm{t}}}$, ${\mathbfcal H}_{u,T}^{{\rm{SFT,d}}}$ can be efficiently generated from ${{\mathbfcal H}^{{\rm{TB}}}_u}$ using FFT. 


\section{Simulation Results}\label{section5}

In this section, we present simulation results to validate the effectiveness of our proposed channel acquisition approach for massive MIMO-OFDM systems. We utilize the well-established QuaDRiGa model \cite{jaeckel2014quadriga} to generate the SFT domain channel tensor for each UT, and the scenario is set to ``3GPP${\rm{\_}}$38.901${\rm{\_}}$UMa${\rm{\_}}$NLOS'' \cite{jaeckel2014quadriga}. The essential simulation parameters are summarized in Table \ref{table2}. We evaluate the performance in two distinct scenarios involving $48$ and $300$ UTs, respectively. Following the channel generation, the statistical CSI ${{\mathbfcal W}_u}$ for each $u \in {\cal U}$ is acquired using the method proposed in \cite{8074806}.



\begin{figure}[t!]
\centering
\includegraphics[width =200pt]{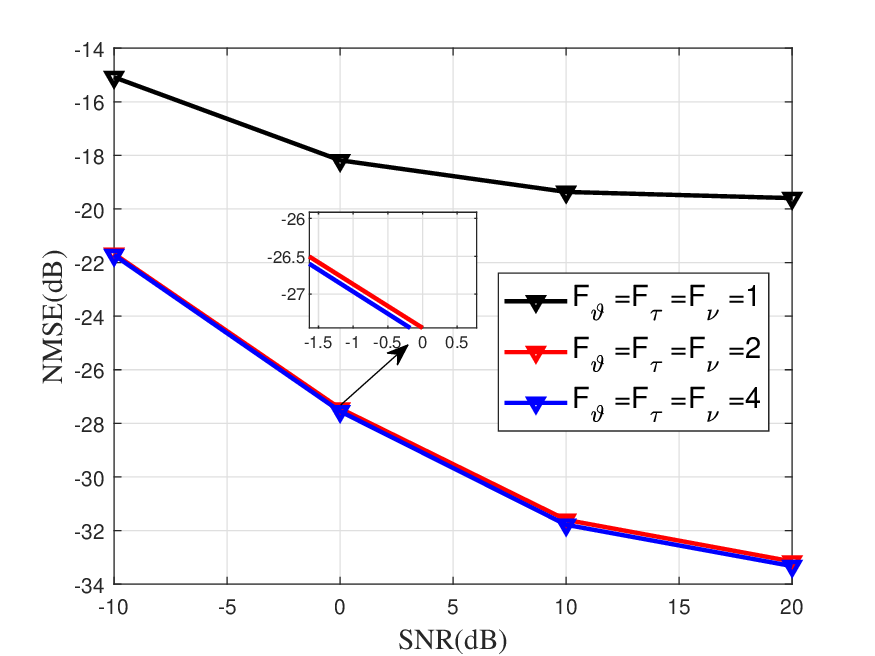}
\caption{NMSE of the proposed channel acquition approach under different fine factor settings, $U=48$, ${v_{{\rm{speed}}}} = 3 {\rm{km/h}}$, ${T_{{\rm{iter}}}} = 300$.}
\label{factor}
\end{figure}

We firstly present the normalized MSE (NMSE) performance of our proposed channel acquisition approach, i.e.,  tensor-based IGA with TFPSPs, under different fine factor settings in Fig. \ref{factor}. The simulation is carried out in the scenario involving 48 UTs at the moving speed ${v_{{\rm{speed}}}}$ = 3 km$/$h, and we consider the case when $T={N_{\rm p}-1}$. The NMSE is defined by
\begin{equation}
\scalebox{1.1}{$\varepsilon _{{\rm{NMSE}}} = \frac{1}{{{N_{{\rm{sam}}}}U}}\sum\limits_{n = 1}^{{N_{{\rm{sam}}}}} {\sum\limits_{u = 0}^{U - 1} {\frac{{\left\| {\hat {\mathbfcal H}_{u,T}^{{\rm{SFT,p}}}\left( n \right) - {\mathbfcal H}_{u,T}^{{\rm{SFT,p}}}\left( n \right)} \right\|_2^2}}{{\left\| {{\mathbfcal H}_{u,T}^{{\rm{SFT,p}}}\left( n \right)} \right\|_2^2}}} },$}
\end{equation}
where ${{N_{{\rm{sam}}}}}$ is the number of channel samples in each OFDM symbol, ${\mathbfcal H}_{u,T}^{{\rm{SFT,p}}}\left( n \right)$ is the $n$-th channel tensor sample of the $u$-th UT and $\hat {\mathbfcal H}_{u,T}^{{\rm{SFT,p}}}\left( n \right)$ is its estimate. It is evident that setting ${F_\upvartheta } = {F_\uptau } = {F_{{\upnu }}} = 2$ results in a significant performance improvement compared to the case where ${F_\upvartheta } = {F_\uptau } = {F_{{\upnu }}} = 1$. At SNR = $20$ dB, setting ${F_\upvartheta } = {F_\uptau } = {F_{{\upnu }}} = 2$ brings a performance gain of around 14 dB compared to ${F_\upvartheta } = {F_\uptau } = {F_{{\upnu }}} = 1$. Moreover, the NMSE with ${F_\upvartheta } = {F_\uptau } = {F_{{\upnu }}} = 4$ is also lower than that with ${F_\upvartheta } = {F_\uptau } = {F_{{\upnu }}} = 2$, despite a small gap between them.

\newcolumntype{L}{>{\hspace*{-\tabcolsep}}l}
\newcolumntype{R}{c<{\hspace*{-\tabcolsep}}}
\definecolor{lightblue}{rgb}{0.93,0.95,1.0}
\begin{table}[H]
\renewcommand\arraystretch{1.2}
\footnotesize
\caption{Parameter Settings}
\centering
\label{table2}
\setlength{\tabcolsep}{0.5mm}{
	\begin{tabular}{LcRcR}
		\toprule
		Parameter & & Value\\
		\midrule
		\rowcolor{lightblue}
		Number of BS antennas, $M$&&$128$&\\
		Number of UTs, $U$&&$48 \& 300$&\\
		\rowcolor{lightblue}
		Carrier frequency $f_{\rm c}$ &&$5.8$ GHz&\\
		Number of subcarriers, $N_{\rm c}$ &&2048&\\	 
		\rowcolor{lightblue}
		CP length $N_{\rm g}$ &&144&\\
		Number of valid subcarriers $K$ &&360&\\	 
		\rowcolor{lightblue}
		Subcarrier spacing  $\Delta f$&&$15$ kHz&\\
		Number of slot in each frame, $N_{\rm p}$ &&8&\\	                 
		\rowcolor{lightblue}
		Number of symbols in each slot, $N_{\rm b}$&&$14$&\\
		Speed of UTs, $v_{\rm {speed}}$ &&3 km$/$h $\&$ 30 km$/$h &\\	                 
		
		
		\bottomrule
\end{tabular}}
\end{table}


We then compare the NMSE performances of different channel acquisition approaches under different UT velocities. We set ${F_\upvartheta } = {F_\uptau } = {F_{{\upnu }}} = 2$ in the folllowing simulations.  The following channel acquisition approaches are compared: 
\begin{itemize}
\item[$\bullet$]
\emph{APSP-IGA}: the channel acquisition is carried out with the SF beam-based channel model \cite{9910031}. APSPs are scheduled based on the SF beam domain statistical CSI \cite{7332961}, and the CSI is then estimated by IGA \cite{9910031,RIGA}.

\item[$\bullet$]
\emph{TFPSP-IGA}: the channel acquisition is carried out with the proposed TB-based channel tensor model. TFPSPs are scheduled by Algorithm \ref{algorithm1}, and the CSI is then estimated by the tensor-based IGA in Algorithm \ref{riga}.

\item[$\bullet$]
\emph{TFPSP-GAMP}: GAMP in \cite{6033942} is used for TB domain channel estimation instead of the tensor-based IGA in TFPSP-IGA.

\item[$\bullet$]
\emph{TFPSP-EPV}: EPV in \cite{9351786} is used for TB domain channel estimation instead of the tensor-based IGA in TFPSP-IGA.


\end{itemize}

\begin{figure}[t!]
\centering
\includegraphics[width =210pt]{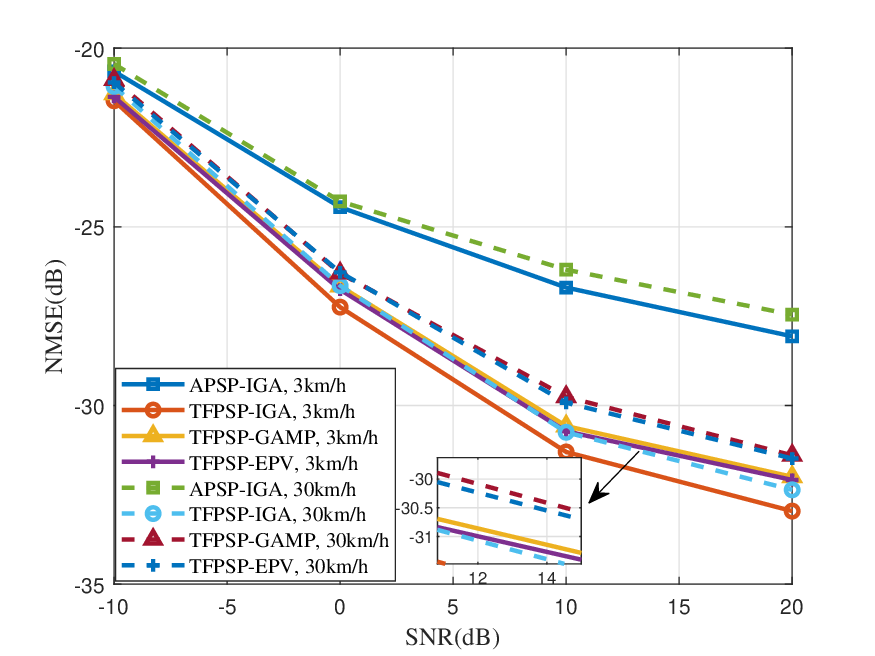}
\caption{NMSE of different channel acquisition approaches versus SNR at different UT speeds, $U=48$, ${T_{{\rm{iter}}}} = 300$.}
\label{mse48}
\end{figure}

The NMSE performances of different channel estimation approaches in the scenario involving $48$ UTs under ${v_{{\rm{speed}}}} = 3{\rm{km/h}}$ and ${v_{{\rm{speed}}}} = 30{\rm{km/h}}$ are shown in Fig. \ref{mse48}, and all the channel estimation approaches are completed within 300 iterations. We use ${T_{{\rm{iter}}}}$ to denote the iteration number. From Fig. \ref{mse48}, it is evident that TFPSP-IGA always delivers a significantly higher performance gain than APSP-IGA. As the SNR increases, the performance gain provided by TFPSP-IGA becomes more pronounced. When SNR = $20$ dB, the NMSE of TFPSP-IGA decreases by approximately 7 dB compared to the NMSE of APSP-IGA. The reason for the aforementioned decrease in NMSE is that, TFPSPs fully exploit the channel sparsity along triple beams and schedule pilots along two dimensions, resulting in lower overlap among the equivalent TB channels and thus achieving better performance. Furthermore, Fig. \ref{mse48} also indicates that, the tensor-based IGA achieves a lower NMSE than GAMP and EPV within the same number of iterations. The SNR gain of the IGA compared to GAMP and VEP is about
1 dB when SNR = $20$ dB.

\begin{figure}[t!]
\centering
\includegraphics[width =210pt]{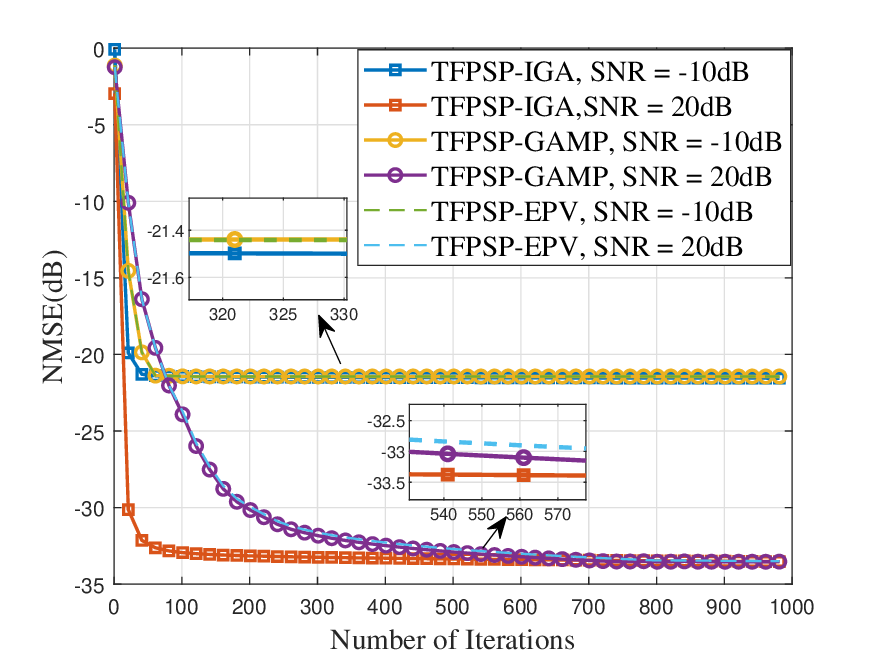}
\caption{Convergence performance versus the number of iterations, $U=48$, ${v_{{\rm{speed}}}} = 3{\rm{km/h}}$.}
\label{iteration}
\end{figure}

Fig. \ref{iteration} illustrates the convergence performances of the tensor-based IGA, GAMP and EPV for SNR set to be $-10$ dB and $20$ dB. It is evident that GAMP and EPV exhibit similar convergence behaviors, while the tensor-based IGA outpaces the other algorithms in convergence speed, particularly in the high SNR scenario. We can observe Fig. \ref{iteration} that, at SNR = 20 dB, IGA requires approximately $180$ iterations to converge, while GAMP and VEP take more than 450 iterations to reach convergence. From \cite{9351786}, it is known that the iteration steps of GAMP and EPV are comparable, involving the update of the first and second moment parameters of the CSI in each iteration. However, according to Algorithm \ref{riga}, in each iteration of the tensor-based IGA, only the parameter ${\mathbfcal L}$ is updated, leading to distinct convergence performance compared to the other algorithms. 


\begin{figure}[t!]
\centering
\includegraphics[width =210pt]{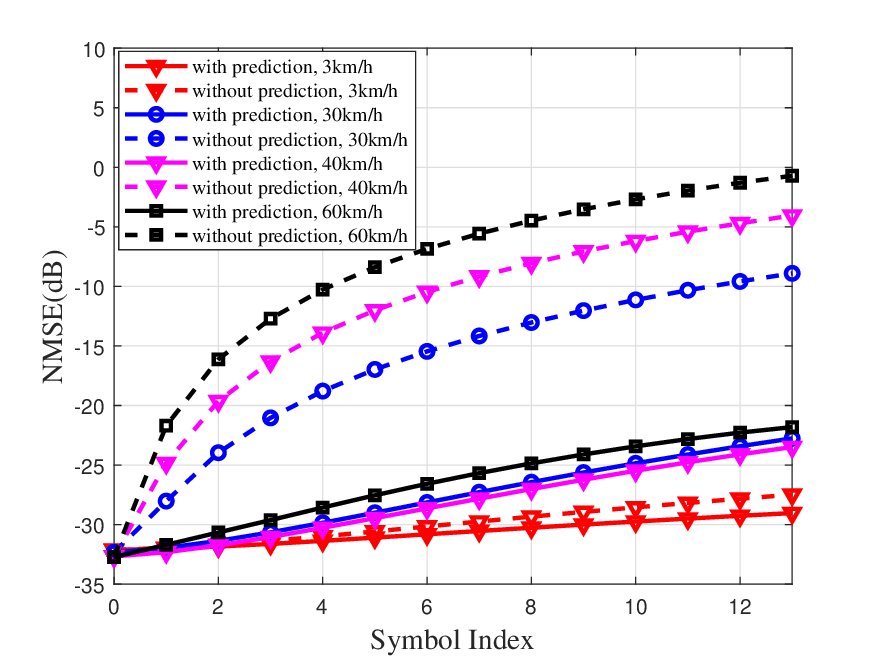}
\caption{NMSE performance versus symbol index at different UT speeds,  $U=48$, SNR = 20 dB.}
\label{prediction}
\end{figure}

To validate the feasibility of the channel prediction method proposed in Section~\ref{section4}, Fig. \ref{prediction} demonstrates the corresponding NMSE performances under different UT speeds. Before the prediction, the CSI in pilot segments is acquired by TFPSP-IGA, where the SNR is set to $20$ dB.  According to Table II, in the current slot, the first OFDM symbol is a pilot segment and the other $13$ OFDM symbols constitute the data segment. Then in this data segment, we implement  channel prediction and calculate the NMSE between the predicted CSI and the actual CSI of each OFDM symbol. In the cases with ${v_{{\rm{speed}}}} = 30{\rm{km/h}}$, ${v_{{\rm{speed}}}} = 40{\rm{km/h}}$ and ${v_{{\rm{speed}}}} = 60{\rm{km/h}}$, our proposed channel prediction provides more precise channel acquisition outcome in comparison to the baseline, where the estimated CSI in the pilot segment is directly applied to the data segments. When ${v_{{\rm{speed}}}} = 3{\rm{km/h}}$, all the UTs move slowly and the CSI changes relatively slow over time. As Fig. \ref{prediction} shows, applying the estimated CSI directly to the data segment can achieve the similar result as channel prediction. By comparing the channel prediction performance at three different speeds, it is evident that as the UT speed increases, CSI varies more rapidly across OFDM symbols, and greater gains can be achieved from channel prediction.


\begin{figure}[t!]
\centering
\includegraphics[width =210pt]{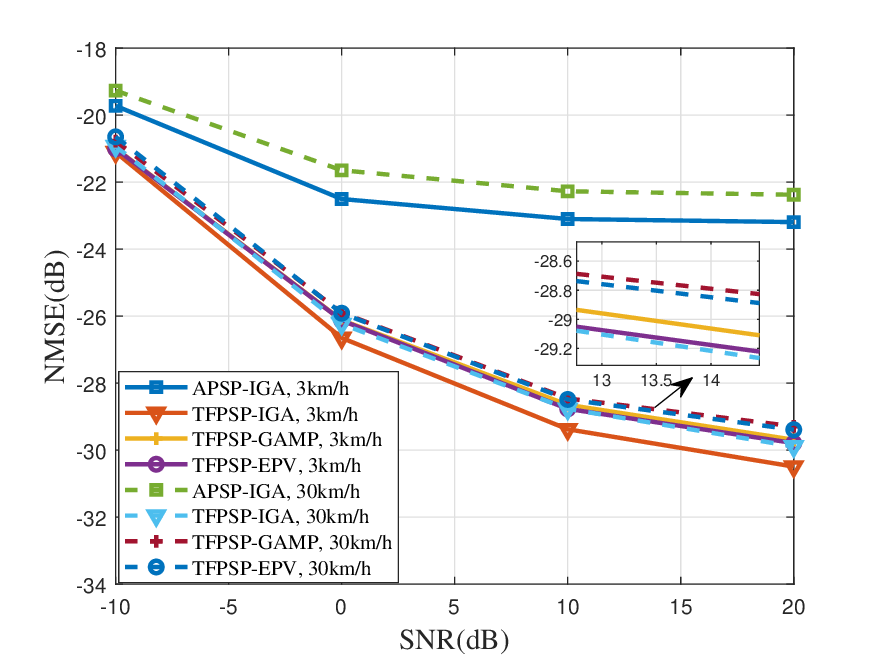}
\caption{NMSE of different channel acquisition approaches versus SNR at different UT speeds, $U=300$, ${T_{{\rm{iter}}}} = 300$.}
\label{mse300}
\end{figure}

We further extend the channel acquisition approaches to a scenario involving 300 UTs. As the number of UTs increases, mitigating interference among them becomes more challenging. The results in Fig. \ref{mse300} show that, compared with APSP-IGA, TFPSP-IGA still yields much more accurate results. As SNR increases, it can be observed that the NMSE curve in Fig. \ref{mse300} corresponding to APSP-IGA exhibits a relatively flat trend. This is because pilot interference among UTs gradually becomes the dominant factor affecting channel estimation performance. In scenarios with a large number of UTs, APSP is unable to effectively suppress inter-UT interference. On the other hand, TFPSP fully exploits the sparse characteristics in the TB domain of the channel and performs pilot scheduling along two dimensions. By scheduling pilots in two dimensions simultaneously, TFPSPs make it easier to mitigate the overlaps among different UTs, enhancing the accuracy of the channel estimation. As a result, especially at high SNR, TFPSP can significantly improve NMSE performance in channel estimation compared to APSP. Fig. \ref{mse300} shows that when the SNR exceeds 10 dB, the NMSE corresponding to TFPSP-IGA decreases obviously compared to APSP-IGA.  When SNR = $20$ dB, the NMSE performance of APSP-IGA is approximately -$22$ dB, while the NMSE performance of TFPSP-IGA is reduced by over 8 dB.  Moreover, within the same number of iterations, i.e., 300 iterations, IGA demonstrates more accurate estimation than GAMP and EPV, especially in high SNR scenarios.
\begin{figure}[t!]
\centering
\includegraphics[width =210pt]{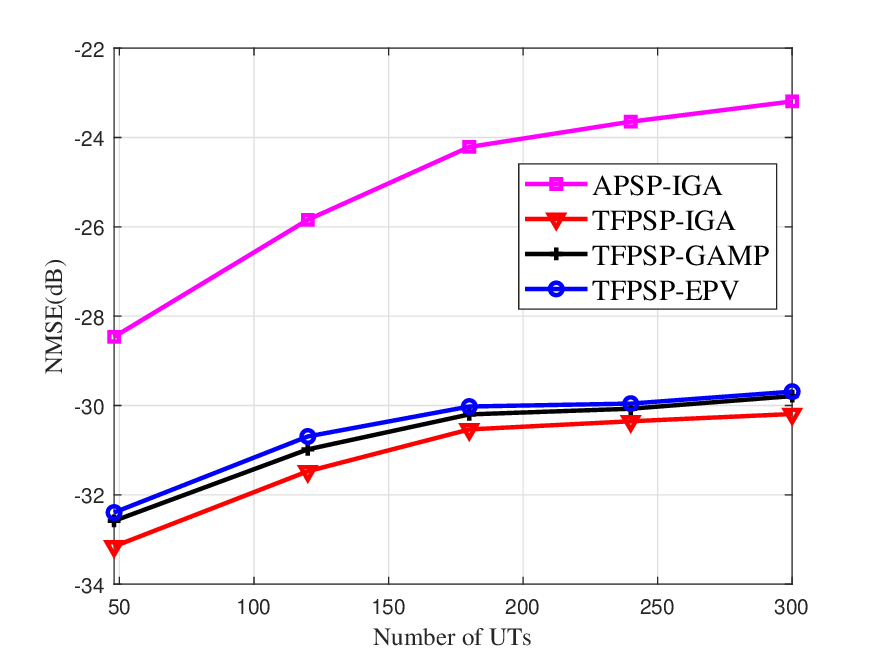}\label{UT111}
\caption{NMSE of different channel acquisition
	approaches versus Number of UTs, ${v_{{\rm{speed}}}}$=3 km$/$h, SNR = 20 dB, and ${T_{{\rm{iter}}}} = 300$.} 
\label{UT}
\end{figure}

To more clearly illustrate the effectiveness of the proposed TFPSPs in suppressing inter-UT interference, we compare the NMSE performance of different channel acquisition methods at SNR = 20 dB and ${v_{{\rm{speed}}}}$= 3 km $/$h, while varying the number of UTs in the system, as shown in Fig. \ref{UT111}. The results clearly demonstrate that the proposed TFPSPs are more effective than conventional FPSPs in suppressing pilot interference, making the NMSE performance of TFPSP-IGA less sensitive to changes in the number of UTs compared to APSP-IGA.

\begin{figure}[t!]
	\centering
	\includegraphics[width =210pt]{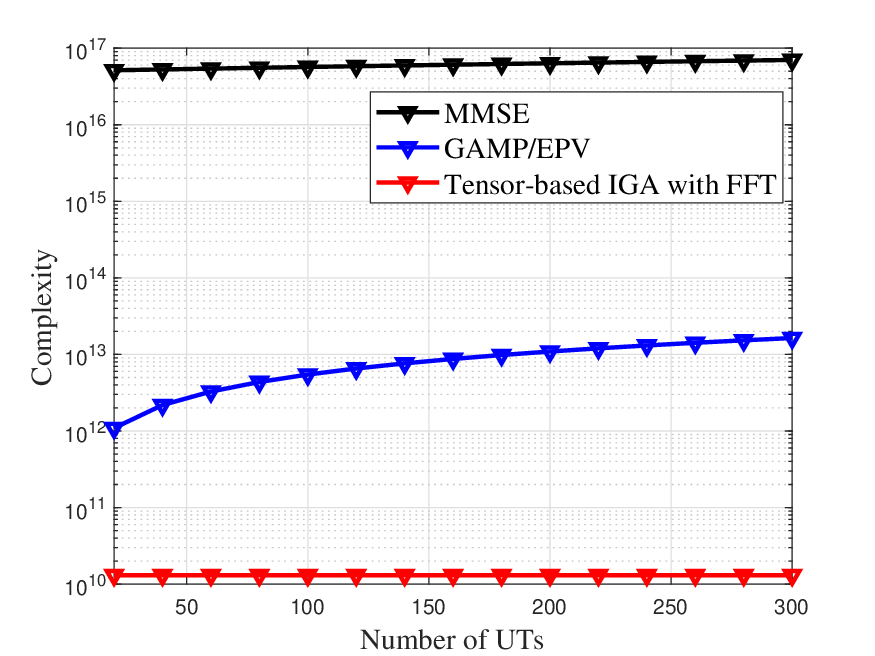}
	\caption{ Complexities of different estimation algorithms, ${T_{{\rm{iter}}}} = 300$.}
	\label{complexity}
\end{figure}

Fig. \ref{complexity} illustrates how the complexities of different algorithms vary with the number of UTs with 300 iterations. From the figure, it is evident that the MMSE algorithm requires the largest complexity due to tensor inversion. In contrast, the complexity of the tensor-based IGA is the lowest, attributed to the full utilization of the specific structure of beam matrices. When $U =300$, the runtimes for 300 iterations of GAMP, EPV, and tensor-based IGA are ${t_{{\rm{GAMP}}}} = 4207$ s, ${t_{{\rm{EPV}}}} = 4119$ s, and ${t_{{\rm{IGA}}}} = 1076$ s, respectively. It is evident that the proposed tensor-based IGA algorithm has a shorter runtime. 

\section{Conclusion}\label{section6}

In this paper, we propose a comprehensive channel acquisition method with TFPSPs for massive MIMO-OFDM systems. To provide a clear understanding of the proposed method, this paper was organized into several key sections: channel modeling, pilot design and scheduling, and channel estimation and prediction. We began by introducing the TB-based channel tensor model, which represents the SFT domain channel as a product of beam matrices and the TB domain channel tensor. Leveraging the unique characteristics of TB domain channels, we developed TFPSPs—a novel two-dimensional pilot design that simultaneously transmits distinct pilot signals in both the time and frequency domains. We then derived the optimal conditions for TFPSPs and proposed a corresponding pilot scheduling algorithm. Additionally, we introduced a tensor-based IGA for estimating TB domain channel tensors, which avoids the need for high-dimensional tensor inversion. By exploiting the structure of beam matrices and the properties of TFPSPs, we developed a low-complexity implementation of the IGA, further reducing the computational burden of channel estimation. Simulation results demonstrated the efficiency of our proposed method, especially in scenarios with a large number of UTs. Although we focus on a single-cell system in this paper, our proposed approach can be extended to more complex systems, including those with more general antenna configurations at both the BS and UT sides, as well as multi-cell systems where pilot contamination needs to be addressed.

\appendices
\section{Proof of Theorem \ref{theorem1}}\label{app1}
First of all, we define ${{\mathbfcal{C}}_u}{\rm{ }} \buildrel \Delta \over = {{{\mathbfcal{R}}_{u,T}^{{\rm{SFT}},{\rm{p}}}}} + \frac{{{\sigma _z}}}{{{\sigma _{\rm{p}}}}}{{\mathbfcal{I}}_{M,{K},{N_{\rm{p}}}}}$. According to the tensor semi-positivity introduced in \cite{opentensor}, ${{\mathbfcal{C}}_{u,{\rm{all}}}^{}} - {\mathbfcal{C}}_u^{} \underline{\succ} {\bf{0}}$ holds, and ${\mathbfcal{C}}_u^{ - 1} - {\mathbfcal{C}}_{u,{\rm{all}}}^{ - 1}\underline{\succ} {\bf{0}}$, leading to 
\begin{equation}
	\scalebox{0.96}{${\rm{tr}}\left\{ {{\mathbfcal R}_{u,T}^{{\rm{SFT}},{\rm{p}}}} \right.\!\!{*_3}{\mathbfcal C}_{u,{\rm{all}}}^{ - 1}{*_3}\!\!\left. {{\mathbfcal R}_{u,T}^{{\rm{SFT}},{\rm{p}}}} \right\} \!\le \!{\rm{tr}}\left\{ {{\mathbfcal R}_{u,T}^{{\rm{SFT}},{\rm{p}}}} \right.\!\!\!{*_3}{\mathbfcal C}_u^{ - 1}{*_3}\!\!\left. {{\mathbfcal R}_{u,T}^{{\rm{SFT}},{\rm{p}}}} \right\}$},
\end{equation}
and $\varepsilon _{{\rm{MSE}}}^{} \ge\varepsilon _{{\rm{MSE}},{\rm{min}}}^{}.$

Then we prove that $\varepsilon _{{\rm{MSE}}}^{} = \varepsilon _{{\rm{MSE,min}}}^{}$ holds when the condition in (\ref{condition}) is satisfied. We calculate the following term 
\begin{equation}\label{RXXR}
	{\mathbfcal R}_{u,T}^{{\rm{SFT}},{\rm{p}}}{*_3}{\mathbfcal R}_{u,u',T}^{{\rm{SFT}},{\rm{p}}} = {\mathbfcal V}_T^{\rm{p}}{*_3}{{\mathbfcal Q}_{u,u'}}{*_3}{\left( {{\mathbfcal V}_T^{\rm{p}}} \right)^{\rm{H}}_3},
\end{equation}
where ${{\mathbfcal Q}_{u,u'}} \buildrel \Delta \over = {\mathbfcal R}_u^{{\rm{TB}}}{*_3}{\left( {{\mathbfcal V}_T^{\rm{p}}} \right)^{\rm{H}}_3}{*_3}{\mathbfcal V}_T^{\rm{p}}{*_3}{\mathbfcal R}_{u,u',T}^{{\rm{TB}}}$. Considering the elements in ${{{\mathbfcal Q}_{u,u'}}}$, we have
\begin{align}\label{de}
	&\scalebox{0.95}{$\quad\quad {\left[ {{{\mathbfcal Q}_{u,u'}}} \right]_{{q_1},{q_2},{q_3},{p_1},{p_2},{p_3}}}$}\vspace{1.5ex}\notag\\
	&\scalebox{0.95}{$ = \displaystyle\sum\limits_{{r_1} = 0}^{M - 1} {\sum\limits_{{r_2} = 0}^{{K} - 1} {\sum\limits_{{r_3} = 0}^{{N_{\rm p}} - 1} {\left\{ {{{\left[ {{\mathbfcal R}_u^{{\rm{TB}}}{ * _3}\left( {{{\mathbfcal V}_T^{\rm{p}}}} \right)_3^{\rm{H}}} \right]}_{{q_1},{q_2},{q_3},{r_1},{r_2},{r_3}}}} \right.} } }$}\notag\\
	&\scalebox{0.95}{$\quad\left. { \cdot {{\left[ {{{\mathbfcal V}_T^{\rm{p}}}{ * _3}{\mathbfcal R}_{u,u'}^{{\rm{TB}}}} \right]}_{{r_1},{r_2},{r_3},{p_1},{p_2},{p_3}}}} \right\}$}\notag\\
	&\scalebox{0.93}{$\mathop {\rm{ = }}\limits^{(a)}   \displaystyle\sum\limits_{{r_1} = 0}^{M - 1} {\sum\limits_{{r_2} = 0}^{{K} - 1} {\sum\limits_{{r_3} = 0}^{{N_{\rm p}} - 1} {\left\{ {{{\left[ {{\mathbfcal R}_u^{{\rm{TB}}}} \right]}_{{q_1},{q_2},{q_3},{q_1},{q_2},{q_3}}} \!\!\!\cdot {{\left[ {\left( {{{\mathbfcal V}_T^{\rm{p}}}} \right)_3^{\rm{H}}} \right]}_{{q_1},{q_2},{q_3},{r_1},{r_2},{r_3}}}} \right.} } } $}\notag \\
	&\scalebox{0.95}{$\quad\cdot{{\left[ {{{{{\mathbfcal V}_T^{\rm{p}}}}}} \right]}_{{r_1},{r_2},{r_3},{p_1},{p_2},{p_3}}}\cdot\left. {{{\left[ {{\mathbfcal R}_{u,u'}^{\rm{TB}}} \right]}_{{p_1},{p_2},{p_3},{p_1},{p_2},p_3^{}}}} \right\}$}\vspace{1.5ex}\notag\\
	&\scalebox{0.95}{$={[{\mathbfcal W}_u^{}]_{{q_1},{q_2},{q_3}}}\cdot{[{\mathbfcal W}_{u',{L_{{\phi _{u'}}}} - {L_{{\phi _u}}},{L_{{\varphi _{u'}}}} - {L_{{\varphi _u}}}}^{}]_{{p_1},{p_2},{p_3}}}$}\notag\\
	&\scalebox{0.9}{$\cdot\underbrace {\sum\limits_{{r_1} = 0}^{M - 1} {\sum\limits_{{r_2} = 0}^{{K} - 1} {\sum\limits_{{r_3} = 0}^{{N_{\rm{p}}} - 1} {\left\{ {{{\left[ {\left( {{{\mathbfcal V}_T^{\rm{p}}}} \right)_3^{\rm{H}}} \right]}_{{q_1},{q_2},{q_3},{r_1},{r_2},{r_3}}} \cdot {{\left[ {{{\mathbfcal V}_T^{\rm{p}}}} \right]}_{{r_1},{r_2},{r_3},{p_1},{p_2},{p_3}}}} \right\}} } } }_{ \buildrel \Delta \over = \alpha \left( {{q_1},{q_2},{q_3},{p_1},{p_2},{p_3}} \right)}$},
\end{align}
where step (a) is because ${{\mathbfcal R}_u^{{\rm{TB}}}}$ is pseudo-diagonal.
From the definition of ${{{\mathbfcal V}_T^{\rm{p}}}}$, ${\alpha \left( {{q_1},{q_2},{q_3},{p_1},{p_2},{p_3}} \right)}$ in (\ref{de}) can be expressed as 
\begin{align}\label{al}
	&\scalebox{0.95}{$\quad\alpha \left( {{q_1},{q_2},{q_3},{p_1},{p_2},{p_3}} \right) $}\notag\\
	&\scalebox{0.95}{$=\displaystyle\sum\limits_{{r_1} = 0}^{M - 1} {\sum\limits_{{r_2} = 0}^{{K} - 1} {\sum\limits_{{r_3} = 0}^{{N_{\rm{p}}} - 1} {\left\{ {\left[ {{{\bf{V}}^{\rm{s}}}} \right]_{{r_1},{q_1}}^*\left[ {{{\bf{V}}^{\rm{f}}}} \right]_{{r_2},{q_2}}^*\left[ {{{\bf{V}}_T^{{\rm{t,p}}}}} \right]_{{r_3},{q_3}}^*} \right.} } }$}\notag\\
	&\scalebox{0.92}{$\quad\left. {\left[ {{{\bf{V}}^{\rm{s}}}} \right]_{{r_1},{p_1}}^{}\left[ {{{\bf{V}}^{\rm{f}}}} \right]_{{r_2},{p_2}}^{}\left[ {{{\bf{V}}_T^{{\rm{t,p}}}}} \right]_{{r_3},{p_3}}^{}} \right\}$}\notag\\
	&\scalebox{0.9}{$=\displaystyle\sum\limits_{{r_1} = 0}^{M - 1} {\sum\limits_{{r_2} = 0}^{{K} - 1} {\sum\limits_{{r_3} = {n_T}}^{T}\!\!\! {\left. {\left\{ {{e^{ - \bar \jmath 2\pi {r_1}\frac{{({p_1} - {q_1})}}{{{N_\upvartheta }}}}}\!\!\!\cdot{e^{ - \bar \jmath 2\pi ({r_2}+{k_0})\frac{{\left( {{p_2} - {q_2}} \right)}}{{{N_\uptau }}}}}\!\!\!\cdot{e^{\bar \jmath 2\pi {r_3}\frac{{ ({{q_3} - {p_3}}) }}{{{N_\upnu }}}}}} \right.} \right\}} } } $}\notag\\
	&\scalebox{0.96}{$=\underbrace {{e^{ - \bar \jmath 2\pi {k_0}\frac{{\left( {{p_2} - {q_2}} \right)}}{{{N_\uptau }}}}} \cdot {e^{ - \bar \jmath 2\pi {n_T}\frac{{\left( {{q_3} - {p_3}} \right)}}{{{N_\upnu }}}}}}_{ \buildrel \Delta \over = \beta_T } \cdot {\alpha _M}\left( {\frac{{{p_1} - {q_1}}}{{{F_\upnu }}}} \right)$}\notag\\
	&\scalebox{0.96}{$\quad\cdot{\alpha _{{K}}}\left( {\frac{{{p_2} - {q_2}}}{{{F_\uptau }}}} \right)\cdot{\alpha _{{N_{\rm{p}}}}}\left( {\frac{{{{q_3} - {p_3}} }}{{{F_\upnu }}}} \right)$},
\end{align}where the function ${\alpha _A}(x) \buildrel \Delta \over = \sum\limits_{a = 0}^{A - 1} {\exp \left( {{\rm{ - }}\bar \jmath 2\pi \frac{ax}{A}} \right)} $$ \to $$\delta \left( x \right)$ as  $A \to \infty $. Substituting (\ref{al}) into (\ref{de}), we have
\begin{align}\label{de2}
	\scalebox{0.9}{${\left[ {{{\mathbfcal Q}_{u,u'}}} \right]_{{q_1},{q_2},{q_3},{p_1},{p_2},{p_3}}}\!={\beta_T\cdot \alpha _M}\left( {\frac{{ {{p_1} - {q_1}} }}{{{F_\upvartheta }}}} \right) {\alpha _{{K}}}\left( {\frac{{{p_2} - {q_2}}}{{{F_\uptau }}}} \right) {\alpha _{{N_{\rm{p}}}}}\left( {\frac{{ {{q_3} - {p_3}} }}{{{F_\upnu }}}} \right)$}\notag\\
	\scalebox{0.95}{$\cdot{[{\mathbfcal W}_u^{}]_{{q_1},{q_2},{q_3}}}\cdot{[{\mathbfcal W}_{u',{L_{{\phi _{u'}}}} - {L_{{\phi _u}}},{L_{{\varphi _{u'}}}} - {L_{{\varphi _u}}}}^{}]_{{p_1},{p_2},{p_3}}}.$}
\end{align}
When the condition in (\ref{condition}) holds,  it is equivalent to  
\begin{equation}\label{www}
	{{ {\mathbfcal W}}_u} \odot {{{\mathbfcal W}_{u',{L_{\phi _{u'}}} - {L_{\phi _u}},{L_{\varphi _{u'}}} - {L_{\varphi _u}}}^{}}} = {\bf{0}}.
\end{equation}
With (\ref{www}), when $p_1=q_1$, $p_2=q_2$ and $p_3=q_3$, ${\big[ {{{ {\mathbfcal W}}_u}} \big]_{{q_1},{q_2},{q_3}}} \cdot {\big[   {{{\mathbfcal W}_{u',{L_{\phi _{u'}}} - {L_{\phi _u}},{L_{\varphi _{u'}}} - {L_{\varphi _u}}}^{}}}      \big]_{{p_1},{p_2},{p_3}}} = 0$ and ${\left[ {{{\mathbfcal Q}_{u,u'}}} \right]_{{q_1},{q_2},{q_3},{p_1},{p_2},{p_3}}} = 0$. Otherwise, when at least one of $p_1\ne q_1$, $p_2\ne q_2$ and $p_3 \ne q_3$ holds,  at least one of  ${\alpha _M}\left(  \cdot  \right)$, ${\alpha _{{K}}}\left(  \cdot  \right)$ and ${\alpha _{{N_{\rm{p}}}}}\left(  \cdot  \right)$ in (\ref{de2}) equals to $0$ as $M,{K},{N_{\rm{p}}}$ $ \to$$ \infty $, which also leads to $\mathop {\lim }\limits_{M,{K},{N_{\rm{p}}} \to \infty } {\left[ {{{\mathbfcal Q}_{u,u'}}} \right]_{{q_1},{q_2},{q_3},{p_1},{p_2},{p_3}}} = 0$. Therefore, the condition in (\ref{condition}) guarantees $\mathop {\lim }\limits_{M,{K},{N_{\rm{p}}} \to \infty } $ ${ {{{\mathbfcal Q}_{u,u'}}} }$ $  = {\bf{0}}$. Substituting this into (\ref{RXXR}), with $M,{K},{N_{\rm{p}}}$ $ \to$$ \infty $, we have ${\mathbfcal R}_{u,T}^{{\rm{SFT,p}}}{ * _3}{\mathbfcal R}_{u,u',T}^{{\rm{SFT,p}}} = {\bf{0}}$. Meanwhile, since ${{\mathbfcal C}_u}{ * _3}{\mathbfcal R}_{u,T}^{{\rm{SFT,p}}} = {\mathbfcal R}_{u,T}^{{\rm{SFT,p}}}{ * _3}{{\mathbfcal C}_u}$, we can derive
\begin{align}\label{der1}
	\scalebox{0.96}{$\!\!\!\!\!	\mathop {\lim }\limits_{M,{K},{N_{\rm{p}}} \to \infty }\!\!{{\mathbfcal C}_u}{*_3}{\mathbfcal R}_{u,T}^{{\rm{SFT,p}}}$}&\scalebox{0.96}{$\mathop  = \limits^{(a)}\mathop {\lim }\limits_{M,{K},{N_{\rm{p}}} \to \infty }{\mathbfcal R}_{u,T}^{{\rm{SFT,p}}}{*_3}{{\mathbfcal C}_u}$}\notag\\&\scalebox{0.96}{$\mathop  = \limits^{(b)} \mathop {\lim }\limits_{M,{K},{N_{\rm{p}}} \to \infty } {\mathbfcal R}_{u,T}^{{\rm{SFT,p}}}{*_3}{{\mathbfcal C}_{u,{\rm{all}}}}.$}
\end{align}
Then, the following relationship holds
\begin{subequations}\label{derall}
	\begin{align}
		&\scalebox{0.96}{$\!\!\!\!\!\mathop {\lim }\limits_{M,{K},{N_{\rm{p}}} \to \infty }\!\!\! {\mathbfcal C}_u^{ - 1}{ * _3}{\mathbfcal R}_{u,T}^{{\rm{SFT,p}}}$}\scalebox{0.96}{$\mathop  = \limits^{} \mathop {\lim }\limits_{M,{K},{N_{\rm{p}}} \to \infty } \!\!\!{\mathbfcal R}_{u,T}^{{\rm{SFT,p}}}{ * _3}{\mathbfcal C}_u^{ - 1}$}\label{der11}\\
		&\scalebox{0.96}{$\!\!\!\!\!\mathop {\lim }\limits_{M,{K},{N_{\rm{p}}} \to \infty }\!\!\! {\mathbfcal C}_u^{ - 1}{ * _3}{\mathbfcal R}_{u,T}^{{\rm{SFT,p}}}$}\scalebox{0.96}{$\mathop  = \limits^{} \mathop {\lim }\limits_{M,{K},{N_{\rm{p}}} \to \infty }\!\!\!{\mathbfcal R}_{u,T}^{{\rm{SFT,p}}}{ * _3}{\mathbfcal C}_{u,{\rm{all}}}^{ - 1},$}\label{der12}\\
		\!\!\!\!\!\!\scalebox{0.96}{$\Rightarrow$} &\scalebox{0.96}{$ \mathop {\lim }\limits_{M,{K},{N_{\rm{p}}} \to \infty } \!\!\!\!\!\!{\mathbfcal R}_{u,T}^{{\rm{SFT,p}}}{ * _3}$}\scalebox{0.96}{${\mathbfcal C}_u^{ - 1}=\mathop {\lim }\limits_{M,{K},{N_{\rm{p}}} \to \infty }\!\!\!\!\!{\mathbfcal R}_{u,T}^{{\rm{SFT,p}}}{ * _3}{\mathbfcal C}_{u,{\rm{all}}}^{ - 1},$} \label{der13}	
	\end{align}
\end{subequations}
where (\ref{der11}) comes from (a) in (\ref{der1}), and (\ref{der12}) comes from (b) in (\ref{der1}). Substituting (\ref{der13}) into the expression of (\ref{NMSE2}), we have ${\varepsilon _{{\rm{MSE}}}} = {\varepsilon _{{\rm{MSE,min}}}}$. This completes the proof.



\bibliographystyle{IEEEtran}
\bibliography{FINAL_VERSION}

\end{document}